\documentclass[camera]{jpaper}
\usepackage{colortbl}
\usepackage{mathptmx}
\newcommand{\ignore}[1]{}
\usepackage{fancyhdr}
\usepackage[normalem]{ulem}
\usepackage{graphicx}
\usepackage{wrapfig}
\usepackage{comment}
\usepackage[affil-it]{authblk}  
\makeatletter
\renewcommand\AB@affilsepx{\hspace{30pt}\protect\Affilfont}
\makeatother

\usepackage{amsmath}
\usepackage{subfig}
\usepackage{color}
\usepackage{xspace}
\usepackage{dblfloatfix}
\usepackage{balance}
\usepackage{hhline}
\usepackage{url}
\usepackage[nocompress]{cite}

\newcommand{\changes}[1]{{\color{black}#1}\xspace}
\newcommand{\changesii}[1]{{\color{black}#1}\xspace}

\title{Techniques for Efficiently Handling Power Surges in Fuel Cell Powered Data Centers: Modeling, Analysis, Results}

\author[$\dagger$]{Yang Li}
\author[$\ast$]{Di Wang}
\author[$\dagger$]{Saugata Ghose}
\author[$\ast$]{Jie Liu}
\author[$\ast$]{Sriram Govindan}
\author[$\ast$]{Sean James}
\author[$\ast$]{\\Eric Peterson}
\author[$\ast$]{John Siegler}
\author[$\dagger$]{Rachata Ausavarungnirun}
\author[$\ddagger$ $\dagger$]{Onur Mutlu}
\affil[$\dagger$]{Carnegie Mellon University} 
\affil[$\ast$]{Microsoft Corporation}
\affil[$\ddagger$]{ETH Z{\"u}rich}

\begin{document}
\maketitle
\pagestyle{plain}
\renewcommand{\headrulewidth}{1pt}
  \fancyhead[C]{\normalsize{SAFARI Technical Report No. 2018-003}}
\baselineskip=11pt
\widowpenalty10000
\clubpenalty10000


\begin{abstract}
Fuel cells are a promising power source for future data centers, offering
high energy efficiency, low greenhouse gas emissions, and high reliability.
However, due to mechanical limitations related to fuel delivery, fuel cells
are slow to adjust to sudden increases in data center power demands, which
can result in temporary \emph{power shortfalls}.  To mitigate the impact of
power shortfalls, prior work has proposed to either perform \emph{power 
capping} by throttling the servers, or \changes{to leverage} \emph{energy storage
devices} (ESDs) that can temporarily provide enough power to make up for the
shortfall while the fuel cells ramp up power generation.  Both approaches have
disadvantages: power capping conservatively limits server performance and can
lead to service level agreement (SLA) violations, while ESD-only solutions must
significantly overprovision the energy storage device capacity to tolerate the shortfalls
caused by \changes{the} \emph{worst-case} (i.e., largest) power surges, which greatly increases the total cost of
ownership (TCO).

We propose SizeCap, the first ESD sizing framework for fuel cell
powered data centers, which coordinates ESD sizing with power capping to enable a cost-effective solution to power shortfalls in data centers.  SizeCap sizes the ESD just large enough to
cover the majority of power surges, but not the worst-case surges that occur infrequently, to greatly reduce TCO.  It then
uses the smaller capacity ESD \changes{\emph{in conjunction with}} power capping to cover the power
shortfalls caused by the worst-case power surges. As part of our new flexible framework, we
propose multiple power capping policies with different degrees of awareness of
fuel cell and workload behavior, and evaluate their impact on workload performance and ESD size.
Using traces from Microsoft's production data center systems, we demonstrate that SizeCap
significantly reduces the ESD size (by 85\% for a workload with infrequent yet
large power surges, and by 50\% for a workload with frequent power surges)
without violating any SLAs.
\end{abstract}


\section{Introduction}
Data center energy consumption has been growing continuously~\cite{DataCenterEnergy}.  In 2013, data centers in the United States
alone consumed an estimated total annual energy of 91~billion kWh, and this is
expected to grow up to as high as roughly 140~billion kWh/year by 2020~\cite{NRDCReport}.
This growth can cause significant increases in the \emph{total cost of
ownership} (TCO)\changes{, along with} increasingly harmful carbon
emission~\cite{CarbonEmission1}.

Fuel cells are one new power source
technology that has been proposed to power data centers with improved
energy efficiency and reduced greenhouse gas emissions~\cite{ChaoLi_ICAC2014,
ChaoLi_HPCA2013,Li_ESFuelCell,LiZhao_TechnicalReport}.  Fuel cells generate 
power by converting fuel (e.g., hydrogen, natural gas) into electricity through
an electrochemical process. Fuel cells have three major advantages.  
First, they have much greater energy efficiency compared to traditional power sources
as they directly convert chemical energy into electrical energy without the
inefficiencies of combustion.  
A recent prototype demonstrates that using fuel cells to power data centers achieves
46.5\% fuel-source-to-data-center energy efficiency,
while using traditional power sources results in only
32.2\% energy efficiency~\cite{LiZhao_TechnicalReport}.
Second, fuel cells lower carbon dioxide emission
by 49\% over traditional combustion based power plants~\cite{LiZhao_TechnicalReport,
FuelCellClean}.  
Third, fuel cells are highly reliable. Natural gas fuel delivery infrastructure is typically buried underground,
and is \changes{robust to threats such as} severe storms~\cite{LiZhao_TechnicalReport, FCReliability}.
These advantages make fuel cells a promising power
source for future data centers.

Unfortunately, fuel cells have a significant shortcoming compared to traditional
power sources, such as an electric grid based source.  An electric grid based
source can deliver total power that is strictly a function of the power
generation capacity; hence, when power demand/load increases, a traditional
power source can adapt rapidly to that demand by changing its power generation
capacity.
In contrast, fuel cells are limited in how rapidly they can increase their fuel
delivery rate as the power demand/load grows~\cite{LiZhao_TechnicalReport, 
Fuel_Cell_Internal_Storage, mueller2009intrinsic}. As a result, they  
\emph{slowly} increase their power output over time, and eventually match the
desired demand only after several seconds or even minutes~\cite{ChaoLi_HPCA2013,
Li_ESFuelCell,mueller2009intrinsic,Fuel_Cell_Internal_Storage,
Fuel_Cell_Controller_Model}. In other words, fuel cells \changes{exhibit a {\em limited load
following behavior}}.  When a power surge occurs, this unique property of fuel
cells results in a period of time where the fuel cells deliver an insufficient
amount of voltage or power, \changesii{resulting in server damage or shut down,} 
which may lead to data center unavailability. We call this 
phenomenon a {\em power shortfall}. 


Currently, there are \changes{two approaches} to mitigating such
power shortfalls.  The first is to perform \emph{power
capping}, where data center \emph{peak} power consumption is restricted to a threshold
value by throttling server execution. Several variants of power capping have
been proposed for traditional utility grid powered data
centers~\cite{Felter2005,goiri.asplos13,govindan.isca11,govindan.asplos12,kontorinis.isca12,wang.asplos14,DiWang_Sigmetrics2012}.
Power capping imposes soft and/or hard limits to the power consumption of a
data center, but has two limitations.
First, its benefits come at the expense of performance. 
Second, its applicability to fuel cells can be limited.
In \changes{the} case of fuel cells, \emph{it is the ramp rate that requires capping, rather than the peak power, as fuel cells can eventually match the desired peak demand.} 
Existing power capping approaches for peak power consumption therefore may not be directly applied, 
as they may unnecessarily restrict load demand and hence cause both \changes{the} fuel cells and data center to be underutilized. 
The second approach uses \emph{energy storage devices} (ESDs, e.g., batteries, supercapacitors) to make up for short duration power
shortfalls caused by load surges. Unlike power capping, leveraging ESDs 
does not incur any performance penalty, as it
allows a fuel cell to observe the actual load and thus increase the
fuel cell's power output over time. Prior work has
proposed to use local ESDs to construct an \emph{uninterruptible power supply}
(UPS) system for each server rack~\cite{ChaoLi_ICAC2014,ChaoLi_HPCA2013,Li_ESFuelCell,LiZhao_TechnicalReport,burke2007batteries,rodatz2005optimal}.
These works have assumed that the UPSes are \emph{sized large enough} to tolerate
the \emph{worst-case} power surge (i.e.,  the largest change of the server load).

We make a major observation from traces of production data center workloads at Microsoft: 
\emph{the worst-case
power surge happens very infrequently}, and in general, \changes{most power
surges are of a much smaller magnitude and much slower ramp rate, and they last for a much shorter
duration, compared to the worst-case power surge}.
Therefore, for most cases, \emph{ESDs sized for worst-case power surges are
significantly overprovisioned}. Unfortunately, such overprovisioning comes at a very
high cost. For example, every kWh increase in supercapacitor capacity 
costs approximately U.S.~\$20,000~\cite{Supercapacitor_Price}.
With an ESD-only solution that uses worst-case sizing, the TCO increases
significantly to cover these infrequent occurrences~\cite{wang.asplos14,DiWang_Sigmetrics2012}. 
The ESD capacity is also limited by the space available within the data center,
as well as the fact that the ESDs may need capacity for other purposes (e.g.,
handling power outages, and accommodating peak power consumption in
tightly-provisioned energy distribution environments~\cite{govindan.isca11,
kontorinis.isca12,DiWang_Sigmetrics2012,urgaonkar.sigmetrics11}).  With all of
these constraints, it is desirable to minimize the ESD capacity dedicated to 
making up for power shortfalls, by sizing this capacity only for the
\emph{typical case} (i.e., to cover \emph{most} of the shortfalls) and relying on secondary solutions to help cover the
infrequent large surges.

To this end, we propose \emph{SizeCap}, an ESD sizing framework for fuel cell powered data centers.  SizeCap
coordinates ESD sizing with power capping to enable a cost-effective solution to power shortfalls in data centers. In this framework, 
an ESD is sized to tolerate only \emph{typical power surges}, thus minimizing the ESD cost and avoiding
unnecessary overprovisioning. In tandem, for the infrequent cases where the worst-case surge occurs, 
SizeCap employs \emph{power capping} to ensure that the load never exceeds the joint handling capability of the fuel cell and the smaller
ESD. As part of our new flexible framework, we propose and evaluate multiple power capping policies for controlling the \emph{load ramp 
rate}\changes{, each} with different levels of awareness of \changes{both the fuel cell's load following behavior and the workload performance
improvement resulting from additional power}.
We design
both centralized capping policies, which can coordinate load distribution across the entire rack,
and decentralized capping policies, where individual servers use heuristics to control their own
power.

We make the following key contributions in this work:
\begin{itemize}
  \item We perform the first systematic analysis of the energy storage sizing
    problem for fuel cell powered data centers. We analyze the impact of various power 
surge parameters on ESD \changes{size}, and \changes{using} both synthetic and real-world data center traces, we
    demonstrate \changes{significantly} different ESD \changes{size} requirements for typical-case and
    worst-case power surges.

  \item We propose to use smaller-capacity ESDs sized only for \emph{typical}
    power surges, and in conjunction propose several new power capping policies that can
    be used alongside the smaller ESDs
to handle infrequent worst-case power surges in fuel cell powered data centers. 

  \item We demonstrate the feasibility and effectiveness of SizeCap in enabling fuel cells, \changes{which are a good power source when the load has limited transience}, to accommodate 
dynamic workloads. 
We show that SizeCap can significantly reduce the TCO of a fuel cell powered data center 
while \changes{still} satisfying the workload service level agreement (SLA) for the production data center workloads we evaluate. 
For example, we can reduce ESD capacity by 85\% for a workload with infrequent but large power surges, and by 50\%
for a workload with frequent power surges, under a reasonable SLA constraint.
\end{itemize}



\section{Background}
\label{sec:background}

\subsection{Fuel Cells}
\label{sec:background:fuelcells}

Fuel cells are an emerging energy source that directly convert fuel into 
electricity through a chemical reaction.  
Unlike conventional combustion-based power generation, fuel
cells do not require an intermediate energy transformation into heat, and 
are therefore not limited by Carnot cycle efficiency~\cite{CarnotCycle1}. As a result, using
fuel cells provides three benefits. First, the energy efficiency from power source to data center increases significantly, from 32.2\% 
to 46.5\%, compared to traditional power sources~\cite{LiZhao_TechnicalReport}. Second, fuel cells reduce carbon emission
over traditional power plants by 49\%~\cite{LiZhao_TechnicalReport,
FuelCellClean}. Third, the delivery infrastructure for natural gas based fuel 
cells is typically buried underground, making fuel cell based power generation
more robust to exposure to \changes{and} damage from threats like severe storms~\cite{LiZhao_TechnicalReport,
FCReliability}.  
These benefits make fuel cells a
promising solution for powering data centers in the near
future~\cite{ChaoLi_ICAC2014,ChaoLi_HPCA2013,Li_ESFuelCell,
LiZhao_TechnicalReport}.

\begin{figure}[h]
    \centering
    \includegraphics[width=\linewidth]{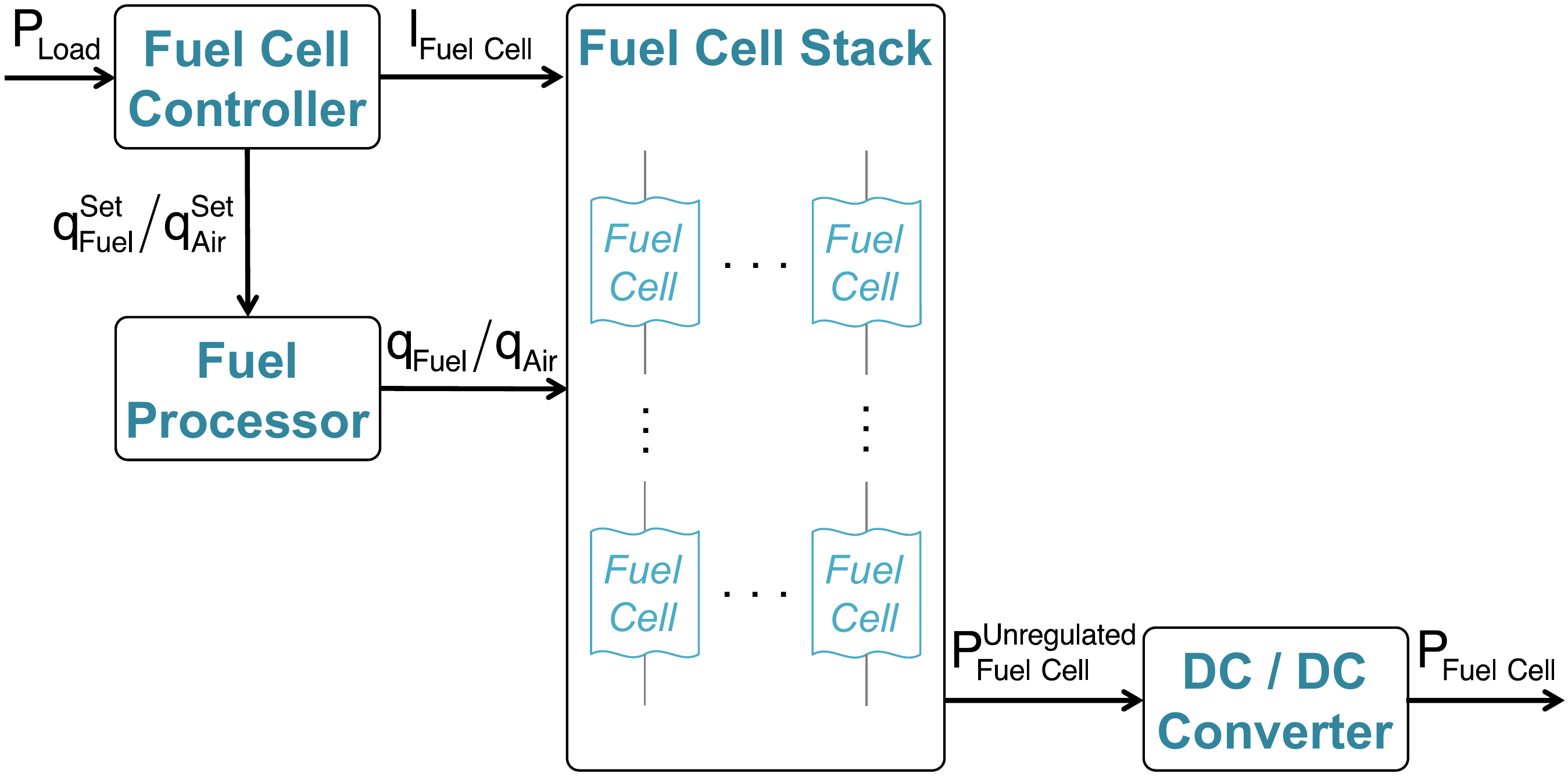}
    \protect\caption{\label{fig:A-typical-fuel}Fuel cell system overview.}
\end{figure}

Figure~\ref{fig:A-typical-fuel} shows the high-level design of a fuel cell
system.  A fuel cell converts the chemical energy
from fuel (e.g., hydrogen, natural gas, biogas) into electricity, typically
through the use of a proton exchange membrane.  Much like batteries, several
fuel cells are combined, both in series and in parallel, to deliver the desired
voltage and current, respectively.  This combined unit, called a \emph{fuel cell
stack}, is managed by a \emph{fuel cell controller}.  The fuel cell controller is an electromechanical
control device that uses the demanded load power ($P_{Load}$) to determine the amount of current that
needs to be generated ($I_{Fuel\: Cell}$), and the desired fuel/air flow rate that needs to be provided ($q_{Fuel}^{Set}$ / $q_{Air}^{Set}$).  The \emph{fuel cell processor} uses this desired fuel/air flow rate to gradually adjust the actual fuel/air flow rate provided for the fuel cell stack ($q_{Fuel}$ / $q_{Air}$) . The fuel cell stack generates power ($P_{Fuel\: Cell}^{Unregulated}$) based on the actual fuel/air flow rate and the fuel cell current that needs to be generated.
The power generated by the fuel cell stack then passes through a DC-to-DC converter before being output ($P_{Fuel\: Cell}$), to
stabilize the output voltage.

A key challenge to directly powering data centers using fuel cells is the limited
\emph{load following} behavior of fuel cells:
a fuel cell incurs a delay before it can fully adjust its
output power to match the demanded load.  
Mechanical limitations in both the fuel cell processor and the fuel delivery system result in a slow response time
to load changes~\cite{LiZhao_TechnicalReport,
Fuel_Cell_Internal_Storage,mueller2009intrinsic}. When the load increases significantly, the
delay due to the limited load following behavior can result in a \emph{power shortfall},
where the fuel cells cannot output enough power to meet the load.  This 
shortfall can cause the servers powered by the fuel cells to crash.

\subsection{Addressing Power Shortfalls with ESDs}
\label{sec:background:esds}

Energy storage devices (ESDs) such as batteries and supercapacitors can be used
to make up for fuel cell power shortfalls. There are two key benefits of ESDs.
First, ESDs can deliver extra power when the data center requires more \emph{power} than
the power generator can provide~\cite{govindan.isca11,kontorinis.isca12, 
DiWang_Sigmetrics2012,urgaonkar.sigmetrics11}. 
Second, ESDs can address power shortfalls: they can be invoked
when there is not enough \emph{current} from the power generator. During
normal operation, fuel cells can generate extra power to recharge ESDs, ensuring
that ESDs are ready to be invoked for future power shortfalls.

However, ESDs come at a cost. First, the size of an ESD has a large impact on
the data center TCO, with supercapacitors today costing approximately
U.S.~\$20,000 per kWh of capacity~\cite{Supercapacitor_Price}. Second, ESDs
take up space in the rack/server. To make matters worse, ESDs will likely 
become multipurpose in the
future~\cite{govindan.isca11,kontorinis.isca12,
DiWang_Sigmetrics2012,urgaonkar.sigmetrics11}, with a single ESD servicing power outages, demand
response, and potentially power shortfalls.  This requires that part of the ESD
capacity be dedicated for these other purposes, further limiting the capacity
available to cover power shortfalls. As a result, while the naive solution is
to size the ESD to cover the largest possible (i.e., \emph{worst-case}) power
shortfall, this may be impractical. In this work, we provide
a solution for data centers to effectively utilize smaller ESDs.

\subsection{Fuel Cell Powered Data Center Setup}
\label{sec:background:datacenter}

\begin{figure}[h]
    \centering
    \includegraphics[width=0.95\linewidth]{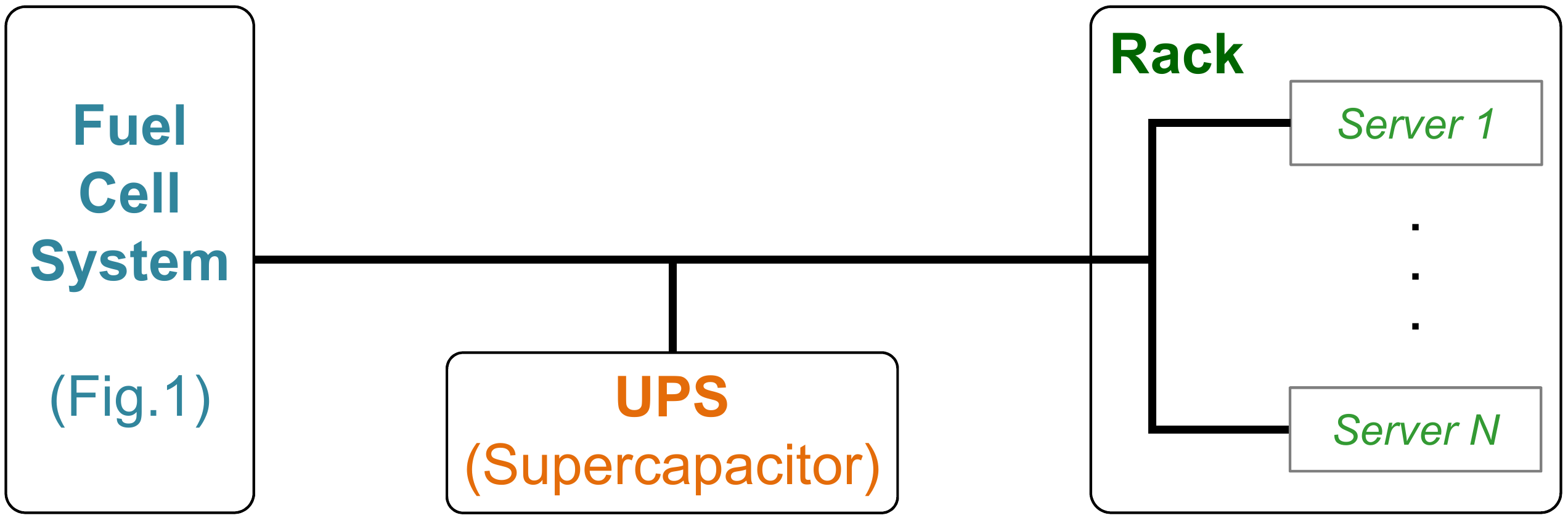}
    \protect\caption{\label{fig:The-system-configuration}Configuration of fuel cell powered data center.}
\end{figure}

In our study, we assume a system configuration as shown in
Figure~\ref{fig:The-system-configuration}.  In this configuration, a fuel cell
system is directly connected with a server rack, and is equipped with a UPS.
Prior prototypes for fuel cell powered data centers have used this same
configuration, as it eliminates unnecessary power equipment (e.g., transformers,
high voltage switching equipment), thereby reducing capital costs and
improving energy efficiency~\cite{Li_ESFuelCell,LiZhao_TechnicalReport}.

We use the ESD within the UPS to cover power shortfalls during spikes in load
power.  We assume that this ESD is a supercapacitor, as when power shortfalls
are no longer than several minutes, supercapacitors are cheaper and more
reliable over their lifetime than batteries~\cite{SchneiderWhitePaper}.  (Prior
work has shown that a fuel cell system can generally match the increased load
power within 5~minutes~\cite{ChaoLi_HPCA2013}.)  We assume that this ESD
discharges whenever there is a power shortfall, and that it is recharged
whenever the fuel cell system matches or exceeds the demanded rack power.
The ESD \changes{subjects} the fuel cell system to a constant load of 1~kW while it
recharges.  We discuss the parameters selected for our data center model in detail in
Section~\ref{sec:methodology}.

\section{Power Shortfall Analysis}
\label{sec:Mot}
\label{sec:analysis}

Fuel cells are slow to react to sudden changes in load power, and exhibit
\emph{load following} behavior, as we discussed in 
Section~\ref{sec:background:fuelcells}.  In order to better understand the
magnitude of the power shortfall problem that results from this behavior, we
analyze the impact of load power changes (i.e., \emph{power surges}) on a data
center powered by fuel cells, with the system configuration shown in 
Figure~\ref{fig:The-system-configuration}.  First, we characterize the extent of
a power shortfall for an example power surge, and then analyze how ESD sizing 
can affect the impact of this shortfall, in Section~\ref{sec:analysis:surges}.
Second, in Section~\ref{sec:analysis:trace}, we study traces from production 
data center systems to determine how ESD size affects availability. 
Finally, in Section~\ref{sec:analysis:capping}, we look at various 
approaches to \emph{power capping}, where servers are throttled to reduce the
upper bound of power consumption when power shortfalls occur.


\subsection{Fuel Cell Reaction to Power Surges}
\label{sub:ESD-Sizing-and}
\label{sec:analysis:surges}

In order to understand the load following behavior of a fuel cell system, we 
study the impact of an example power surge on a detailed model of fuel cell
behavior (see Section~\ref{sec:methodology}). We model the power surge as a sequence of loads
applied over time:
\begin{enumerate}[leftmargin=25pt]
\itemsep 0pt \parskip 0pt
\item 0--2 minutes: constant load of 5.6~kW
\item 2--3 minutes: constant slope ramp up to 10.3~kW
\item 3--12 minutes: constant load of 10.3~kW
\item 12--13 minutes: constant slope ramp down to 5.6~kW
\item 13--15 minutes: constant load of 5.6~kW
\end{enumerate}

\begin{figure}[t]
    \centering
    \includegraphics[width=7.5cm]{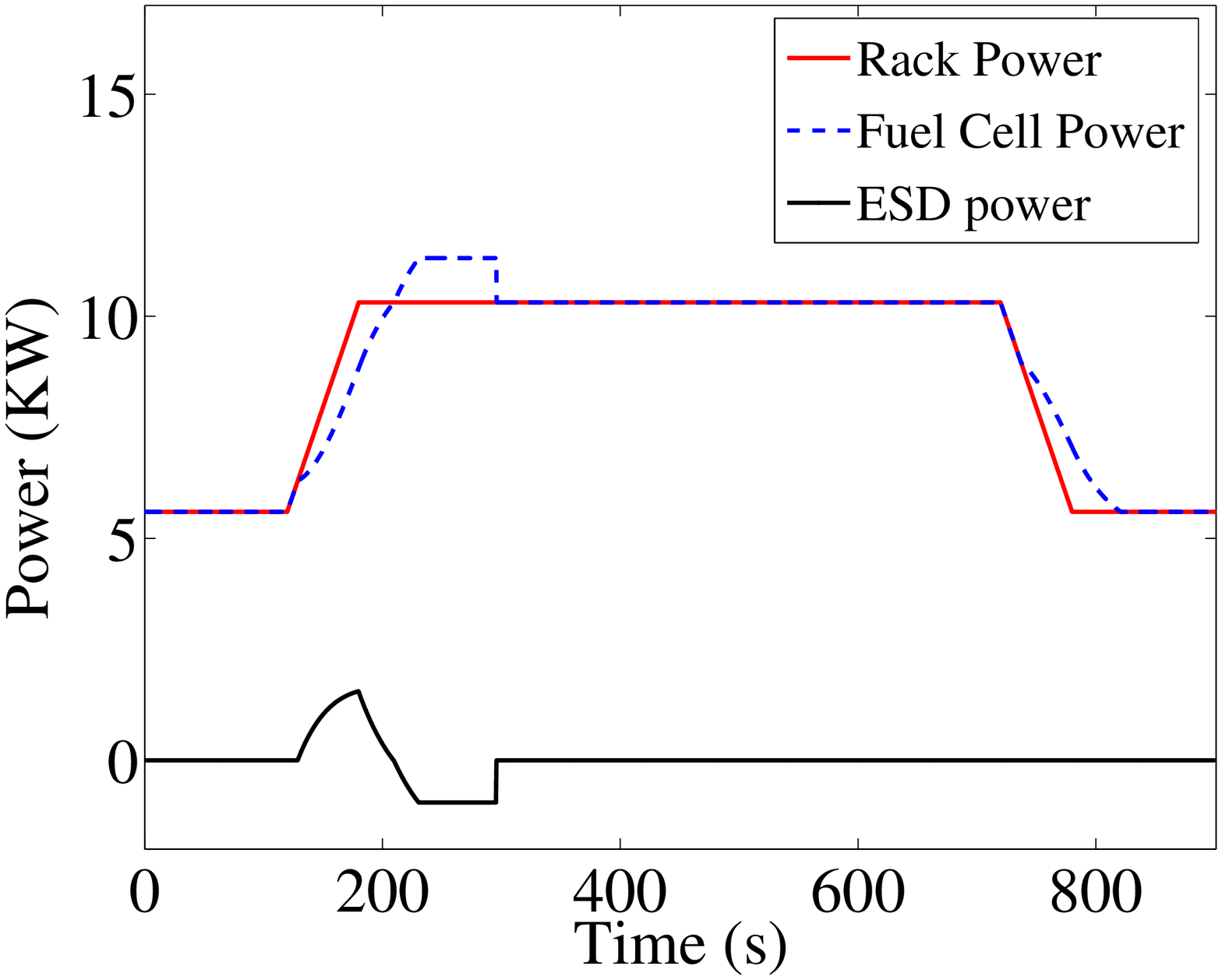}%
    \vspace{-4.5pt}
    \protect\caption{Fuel cell and ESD power output when the load power exhibits a surge.}
    \label{fig:A-typical-power}
    \vspace{-2pt}
\end{figure}

Figure~\ref{fig:A-typical-power} illustrates this load as a solid red line.  We
also plot the output power of the fuel cell system when it is subjected to this
load.  Under our example power surge, the fuel cell system \changes{matches} the load
power 1.5~minutes after the ramp up begins.  When the ramp up begins, we
observe that initially, the fuel cell system can keep up with the increase, as
there is enough fuel already within the fuel cell stack to rapidly increase power production.  
However, once the output power reaches 6.3~kW, the system can continue to
increase power \changes{only} as the fuel flow rate increases, which, as we described in
Section~\ref{sec:background:fuelcells}, is slow.  This slowdown in power output
increase results in a power shortfall, as Figure~\ref{fig:A-typical-power} 
shows.  At this point, the ESD begins to discharge in order to make up for the
shortfall.  After 3.5~minutes have elapsed (i.e., 1.5 minutes after the ramp up
started), the fuel cell power output finally matches the load.  
In order to accommodate this shortfall, we need an ESD with a minimum
capacity of 91~kJ.\footnote{In order to guarantee normal operation of the UPS, we ensure that the stored energy of
the ESD within the UPS never drops below 20\% of its
capacity.  We call this value the \emph{energy threshold}. This constraint is accounted for when we calculate the minimum required ESD capacity.}
\changes{Note that even though the power output now matches the load, 
the fuel cell system briefly continues to increase its output, as it must now 
recharge the ESD in addition to fully powering the rack.}

In order to understand how variation in power surges impacts the capacity
required for the ESD, we characterize these surges using the three properties
shown in Figure~\ref{fig:A-typical-power}: (1)~\changes{\emph{magnitude}} (the difference between
initial power and peak power), (2)~\changes{\emph{slope}} (the rate at which the load ramps up),
and (3)~\changes{\emph{width}} (how long the surge lasts for).  Our initial example power surge
has a magnitude of 4.7~kW, a slope of 78~W/s, and a width of 11~minutes.  We
perform a controlled study of each property, varying the values for one property
\emph{while we hold the other two properties constant at these values}.

\begin{figure}[h]
    \centering
    \vspace{-5pt}
    \includegraphics[width=7cm]{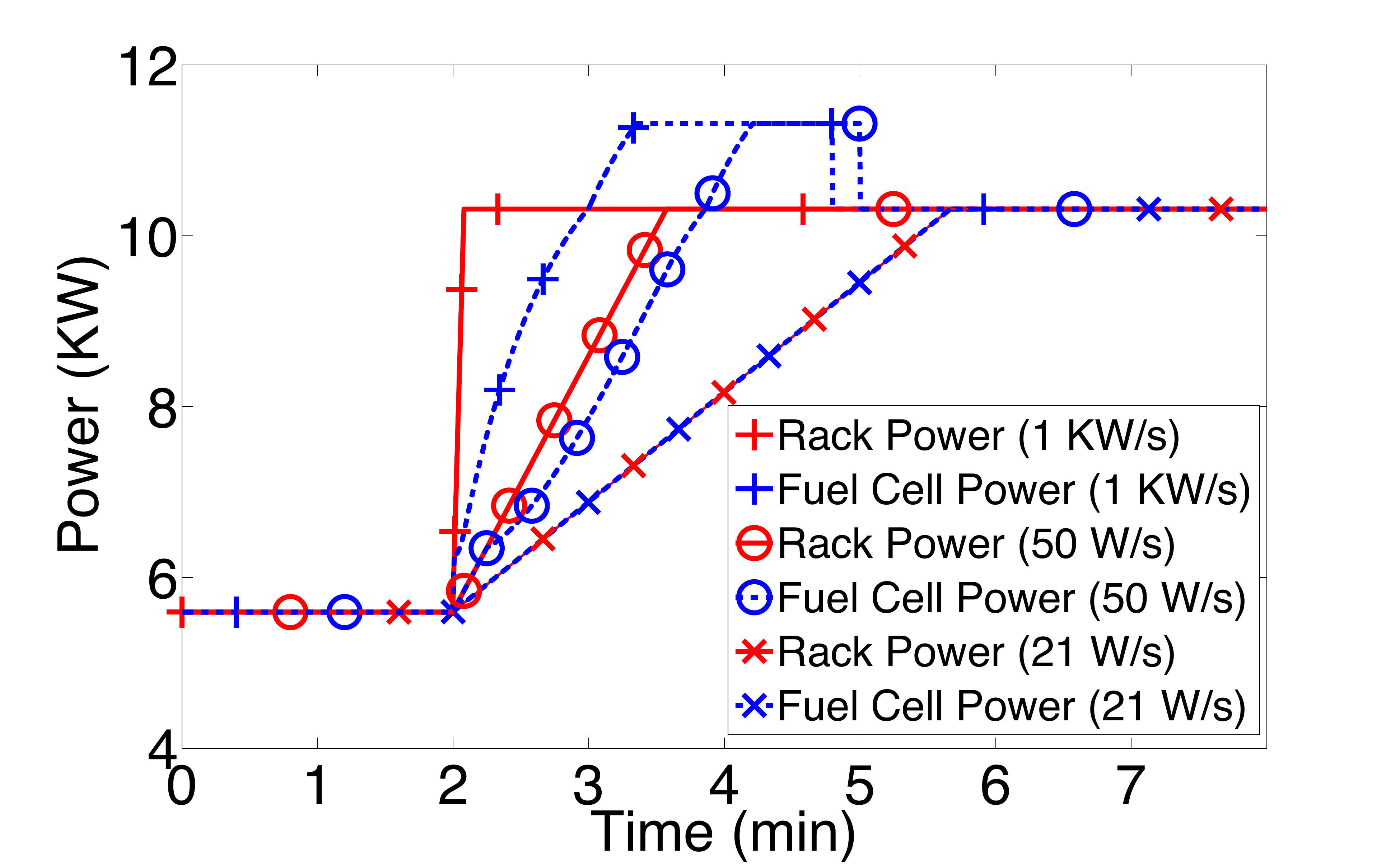}%
    \vspace{-2pt}
    \protect\caption{Fuel cell load following behavior for
different surge slopes during power surge ramp up.}
    \label{fig:slope_example}
    \vspace{-1pt}
\end{figure}

Figure~\ref{fig:slope_example} shows how the fuel cell responds when we vary
the surge \emph{slope}.  For a reduced slope of 21~W/s, we find that the fuel
cell system can always supply the power demanded, and there is no
shortfall.\footnote{Upon further experimentation, we find that for larger
magnitudes, there can be shortfalls with a \changes{surge} slope of 21~W/s.  We find that at a
slope of 16~W/s, the fuel cell system never experiences a shortfall 
within the rack power operating range, 
regardless of the magnitude of the surge.  This slope is defined to be the
\emph{load following rate} of the fuel cell system.}  For a slope of 50~W/s, we see
a slight shortfall, where the fuel cell can increase its power at an 
average rate of only 42~W/s.  A larger slope of 1~kW/s experiences a large shortfall,
with the fuel cell system capable of increasing its output power at an 
average rate of only 78~W/s.  As was the case with our example surge in
Figure~\ref{fig:A-typical-power}, we need an ESD to make up for this shortfall.

\begin{figure*}[t]
\vspace{-8pt}
\subfloat[]{\centering
\includegraphics[width=5.8cm]{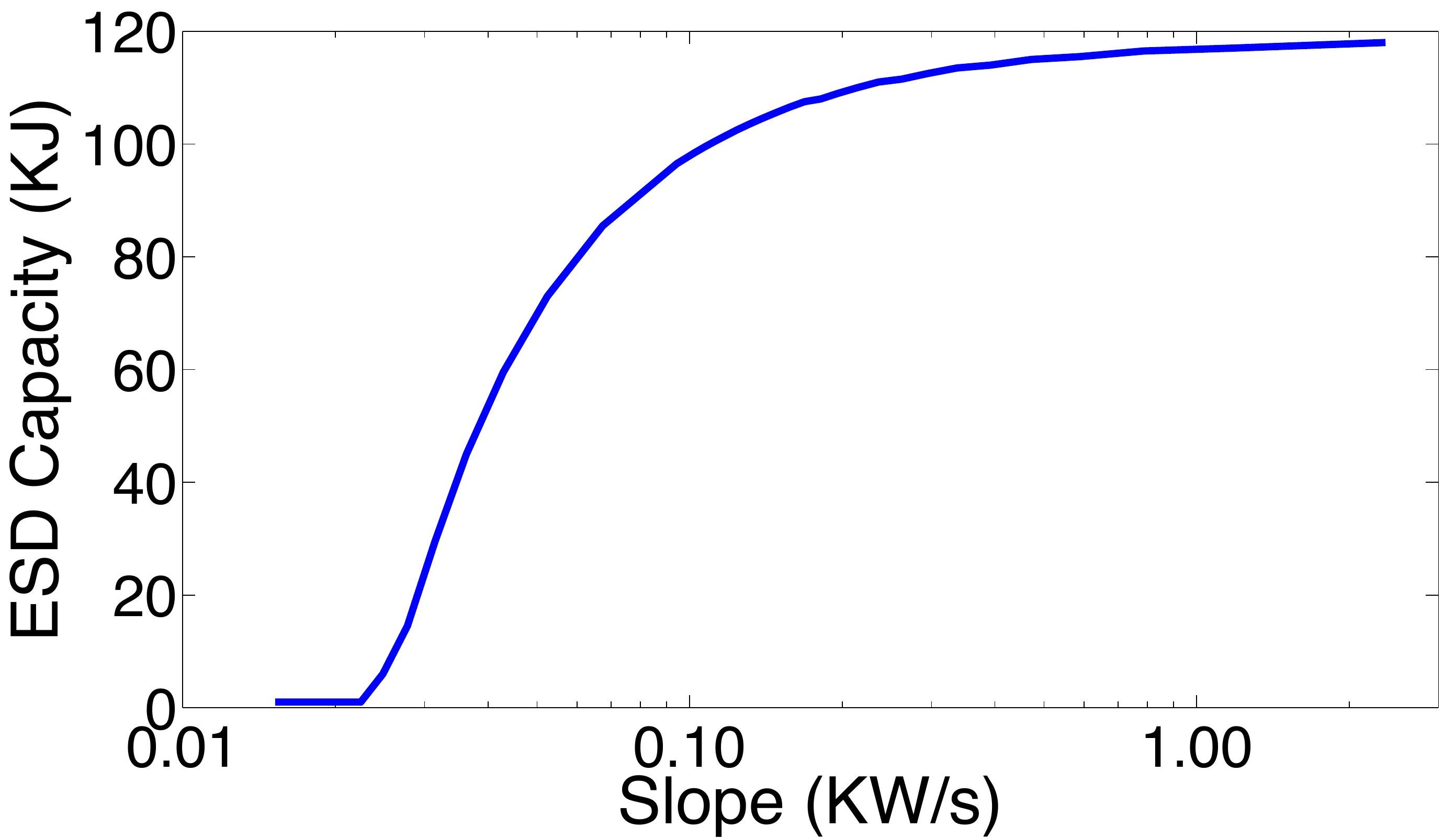}}
\hfill{}
\subfloat[]{\centering
\includegraphics[width=5.65cm]{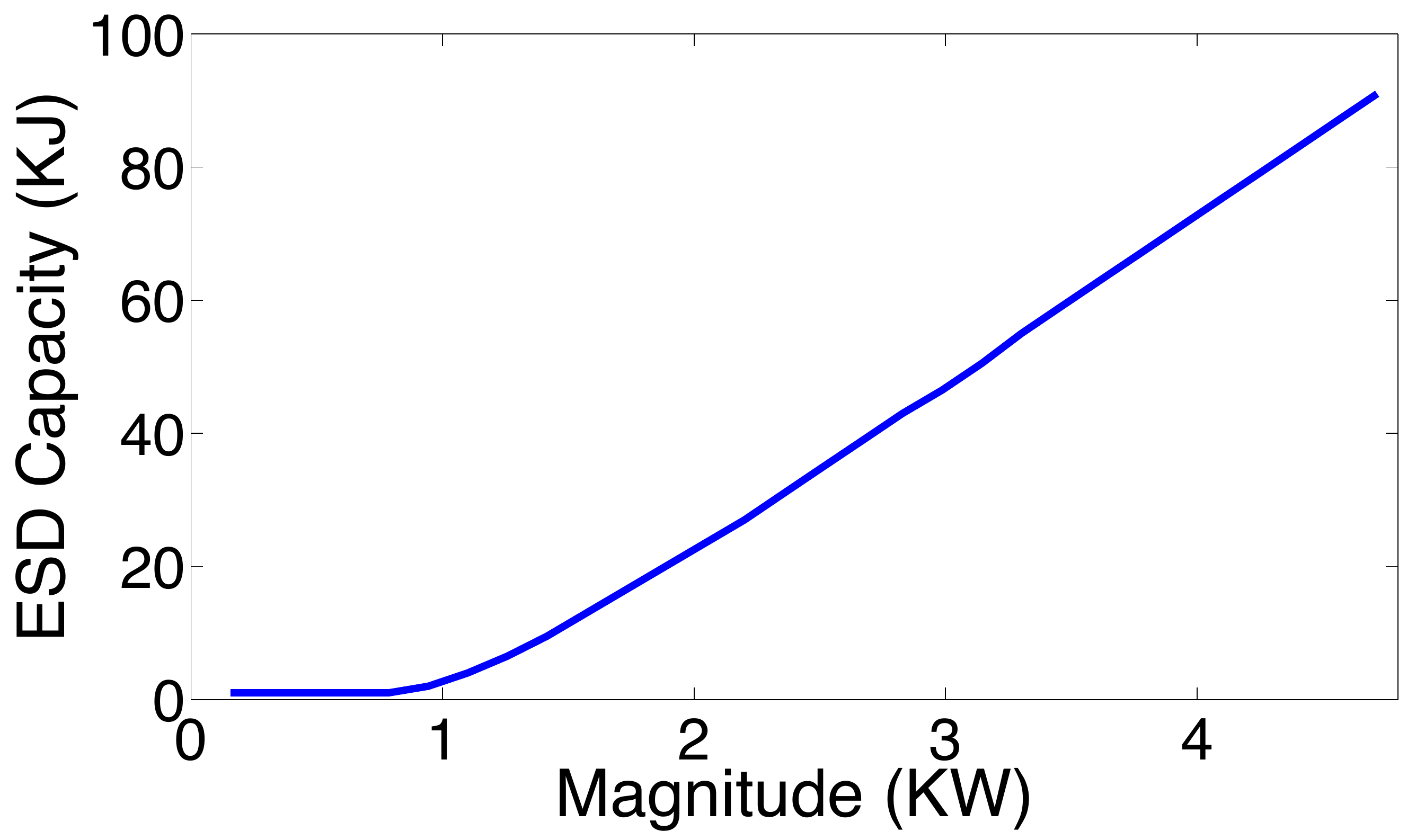}}
\hfill{}
\subfloat[]{\centering
\includegraphics[width=5.8cm]{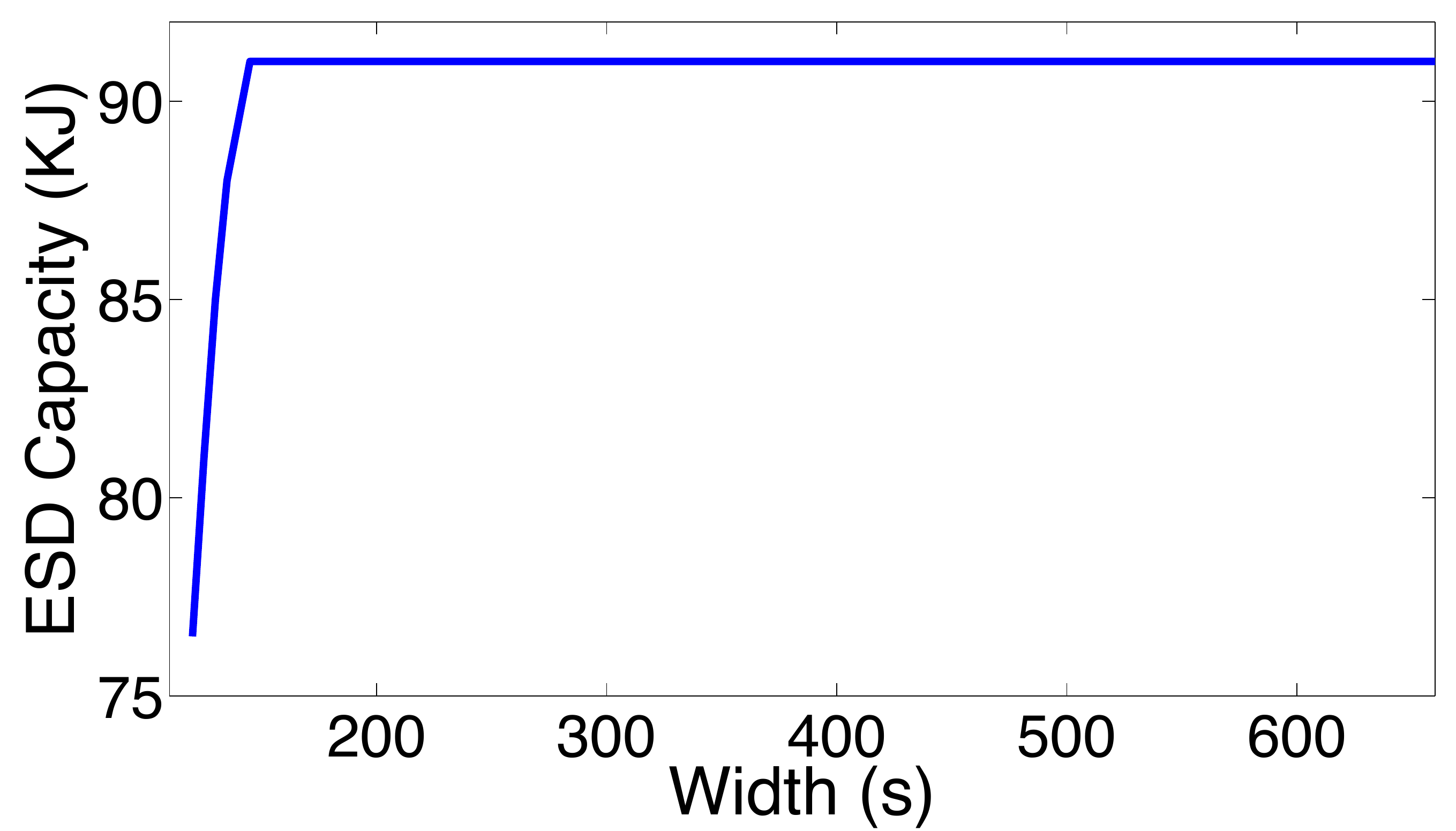}}
\vspace{-3pt}
\protect\caption{Impact of different power surge characteristics
on ESD size: (a) slope, (b) magnitude, and (c) width.}
\label{fig:parameter_impact}
\end{figure*}

As Figure~\ref{fig:slope_example} demonstrates, the amount of power shortfall
increases with the slope of the power surge. We plot the relationship between
surge slope and the required ESD size to handle a power shortfall (again holding magnitude and width constant) in
Figure~\ref{fig:parameter_impact}a.  As we found in Figure~\ref{fig:slope_example},
small slope values do not necessitate an ESD.  We perform similar sweeps to show
the relationship between surge magnitude and ESD size
(Figure~\ref{fig:parameter_impact}b), and between surge width and ESD size
(Figure~\ref{fig:parameter_impact}c).  
We find that when the magnitude is small, we do not
need any ESD. This is because the fuel cell can leverage its internal
fuel to increase its output power in a timely manner and handle the power surge. However, as magnitude continues
to grow, the required ESD capacity needs to grow linearly since the limited internal fuel cannot handle such large
power surges.
Besides, we also find that when the width is short, the required ESD
capacity grows with the width. However, as the width continues
to grow, the required ESD capacity remains constant, since fuel
cell power eventually matches the load power, at which point the ESD is no longer required.

In summary, we find that the slope, magnitude, and width of a power surge are important
characteristics required to determine the minimum ESD capacity required to avoid
a power shortfall.

\subsection{Impact of ESD Size on Availability}
\label{sec:analysis:trace}

\begin{figure*}[t]
\vspace{-5pt}
\subfloat[]{\centering
\includegraphics[width=5.65cm]{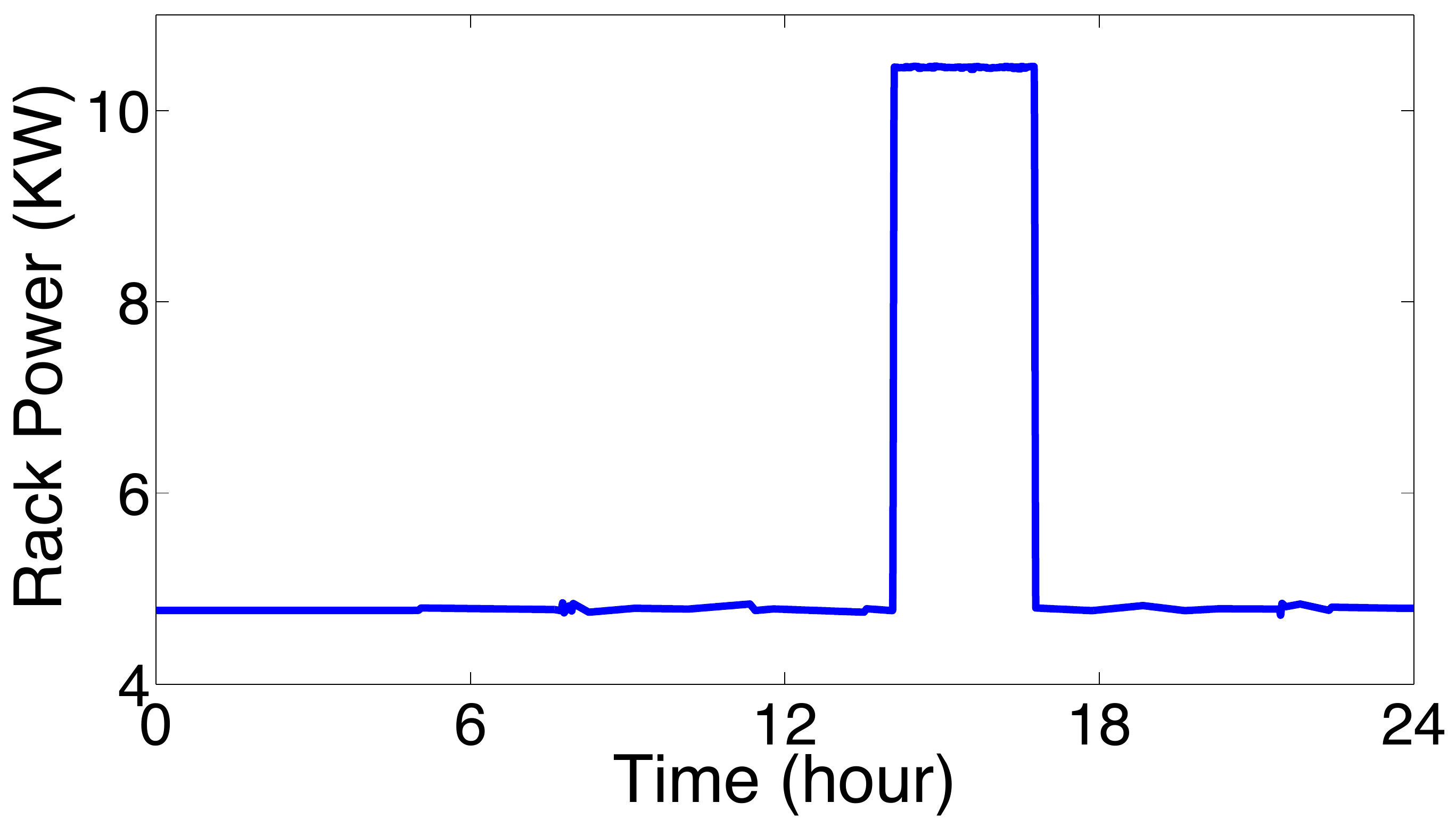}}\hfill{}\subfloat[]{\centering
\includegraphics[width=5.8cm]{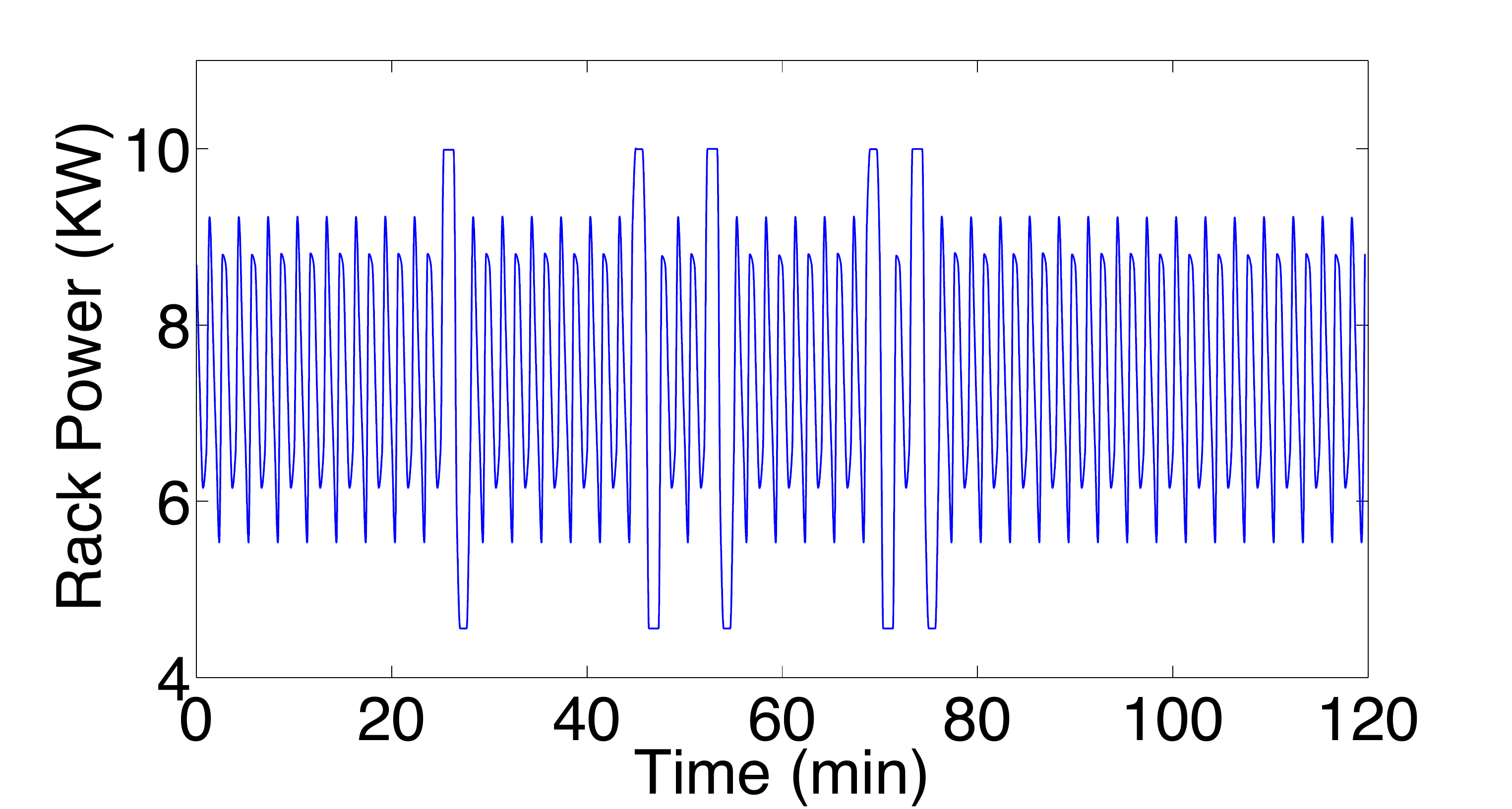}}\hfill{}\subfloat[]{\centering
\includegraphics[width=5.9cm]{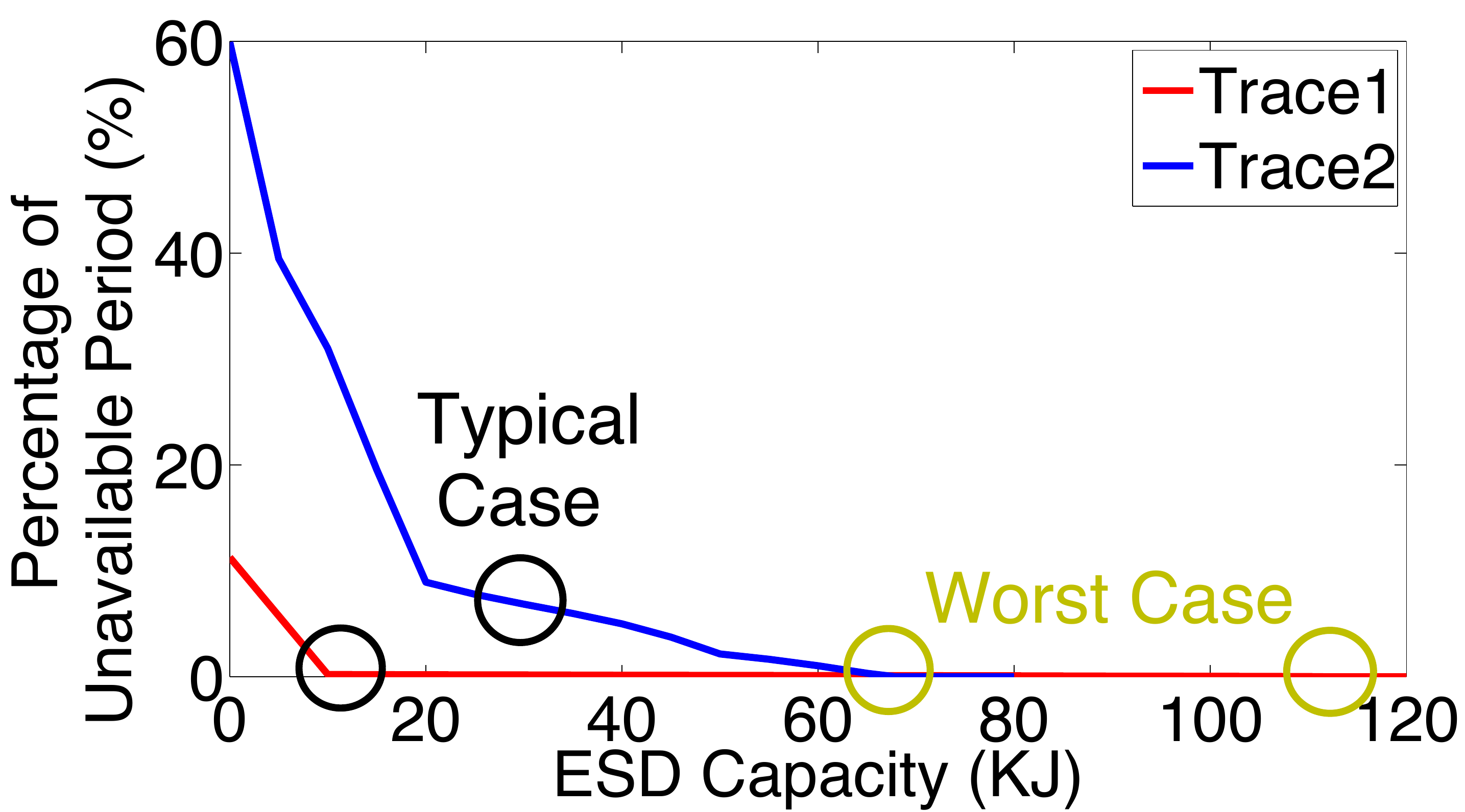}}
\vspace{-2pt}
\protect\caption{\label{fig:Real-world-trace-analysis}Load power traces for production
data center workloads: (a) Trace~1, (b) Trace~2; (c) Data center availability for Traces~1
and~2 when ESD capacity is reduced.}
\end{figure*}

We now investigate the impact that ESD size has on data center availability.
Figures~\ref{fig:Real-world-trace-analysis}a and~\ref{fig:Real-world-trace-analysis}b
show two load power traces recorded from Microsoft's production data centers (see Section~\ref{sec:methodology} for details).
Trace~1 (Figure~\ref{fig:Real-world-trace-analysis}a) is a case where
the rack power demand remains flat for most of the time, except for a single,
large power surge.  Trace~2 (Figure~\ref{fig:Real-world-trace-analysis}b)
is a case that includes several power surges, each with different surge
slopes, magnitudes, and widths.

For both of these traces, we sweep over various ESD sizes to determine the
percentage of time the rack \changes{is} unavailable (i.e., there \changes{is} an
unmitigated power shortfall) for each ESD size, as shown in
Figure~\ref{fig:Real-world-trace-analysis}c.  In order to avoid any shortfalls,
we traditionally size our ESD to handle the \emph{worst-case} power surge (i.e.,
the greatest observed change in load power).  However, we observe that if we
reduce the size of the ESD somewhat, such that it can cover the vast majority
of the power surge behavior (i.e., \emph{typical} power surges), the
data center unavailability remains very low.  For Trace~1, we can cut down the
ESD size by as much as 85\%, while experiencing only 0.4\% unavailability.  For
Trace~2, reducing the ESD size by 50\% leads to unavailability of only 6\%.

We conclude that the ESD size can be reduced significantly if we size it
to cover only \emph{typical} power surges instead of \emph{worst-case} surges.
However, this \changes{introduces} data center unavailability when the
infrequent worst-case surges do occur.  As we discuss next in 
Section~\ref{sec:analysis:capping}, we can employ \emph{power capping} as a secondary power control
mechanism, to ensure that the data center remains available even in
the infrequent cases where the ESD is no longer large enough to cover the shortfall.


\subsection{Employing Power Capping}
\label{sec:analysis:capping}

A second method of preventing power shortfalls is \emph{power capping}, where
the rack servers are throttled to ensure that their power consumption does not
exceed the amount of available power in the system.  On its own, power capping
can be unnecessarily restrictive: power shortfalls in a fuel cell powered data 
center are only temporary until the fuel cell system can ramp up its power
production, but power capping alone prevents the server power consumption from
increasing beyond a certain value, even if the fuel cell can eventually 
deliver that power.  

We instead choose to implement power capping \emph{on top of using an ESD} for shortfall 
mitigation.  Our primary solution to avoiding power shortfalls is to rely on a
\emph{smaller} ESD, which, as we mentioned in Section~\ref{sec:analysis:trace},
can cover the majority of typical power surges.  In the infrequent cases where
the ESD is not large enough to handle worst-case power surges, we propose to use power capping
to ensure that the load power demanded by the rack does not exceed
the combined power output of the fuel cell system and the ESD.

\begin{figure}[t]
    \centering
    \includegraphics[width=7cm]{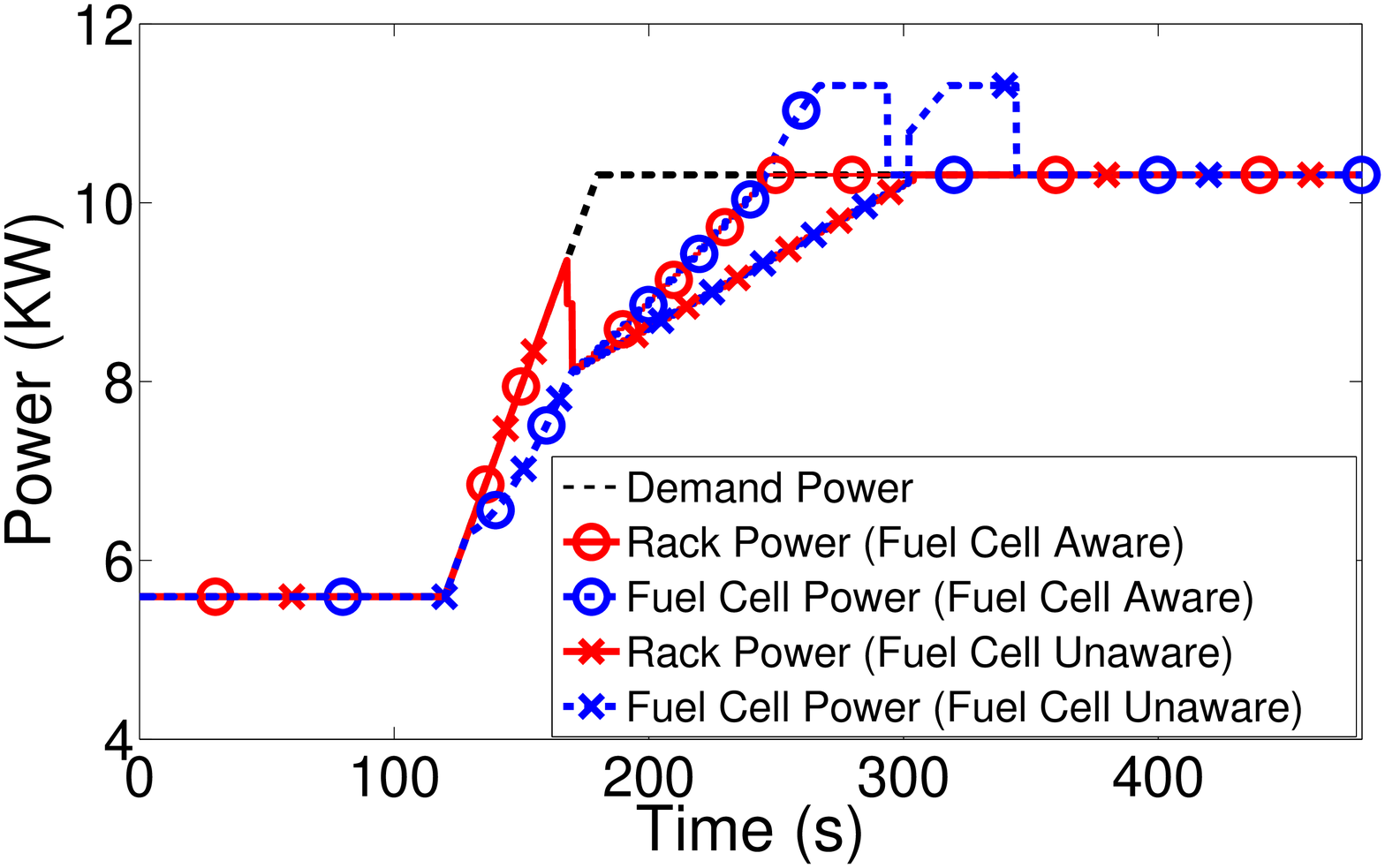}%
    \protect\caption{Behavior of fuel cell aware vs.\ fuel cell unaware power capping policies
    during power surge ramp up.}
    \label{fig:Fuel-cell-aware}
\end{figure}

There are several options for implementing power capping.  One option is to
make the power capping policy \emph{fuel cell aware}.  A fuel cell aware policy
has knowledge of the fuel cell system's load following behavior, and works to
increase the load power as quickly as possible to ensure that the fuel cell ramps
up at its fastest possible rate, whereas a fuel cell unaware policy performs 
more conservative capping.  Figure~\ref{fig:Fuel-cell-aware} shows the two
capping policies being used on our example power surge from 
Section~\ref{sec:analysis:surges}.  As we can see, both policies behave 
identically until around the 170~second mark, at which point the ESD is no longer
able to make up for the shortfall.  We see that the fuel cell aware capping
policy allows the fuel cell to ramp up to full power output much earlier, with
an average power increase of 29~W/s.  In contrast, fuel cell unaware capping
restricts the power output ramp up to only 16~W/s.

\begin{figure}[t]
\vspace{-8pt}
\subfloat[]{
\centering
\includegraphics[width=4cm]{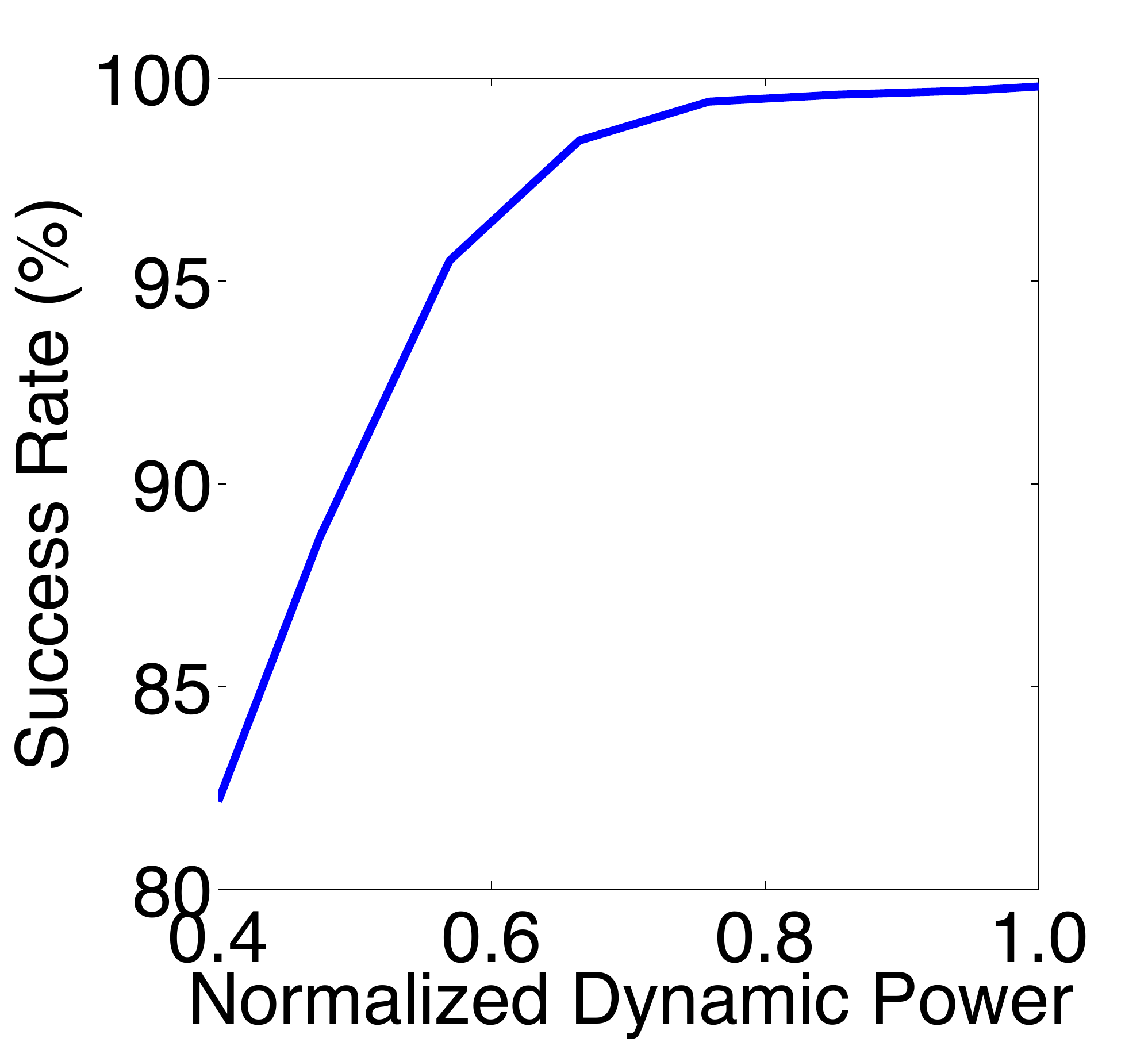}
}\hfill{}\subfloat[]{\centering
\includegraphics[width=4cm]{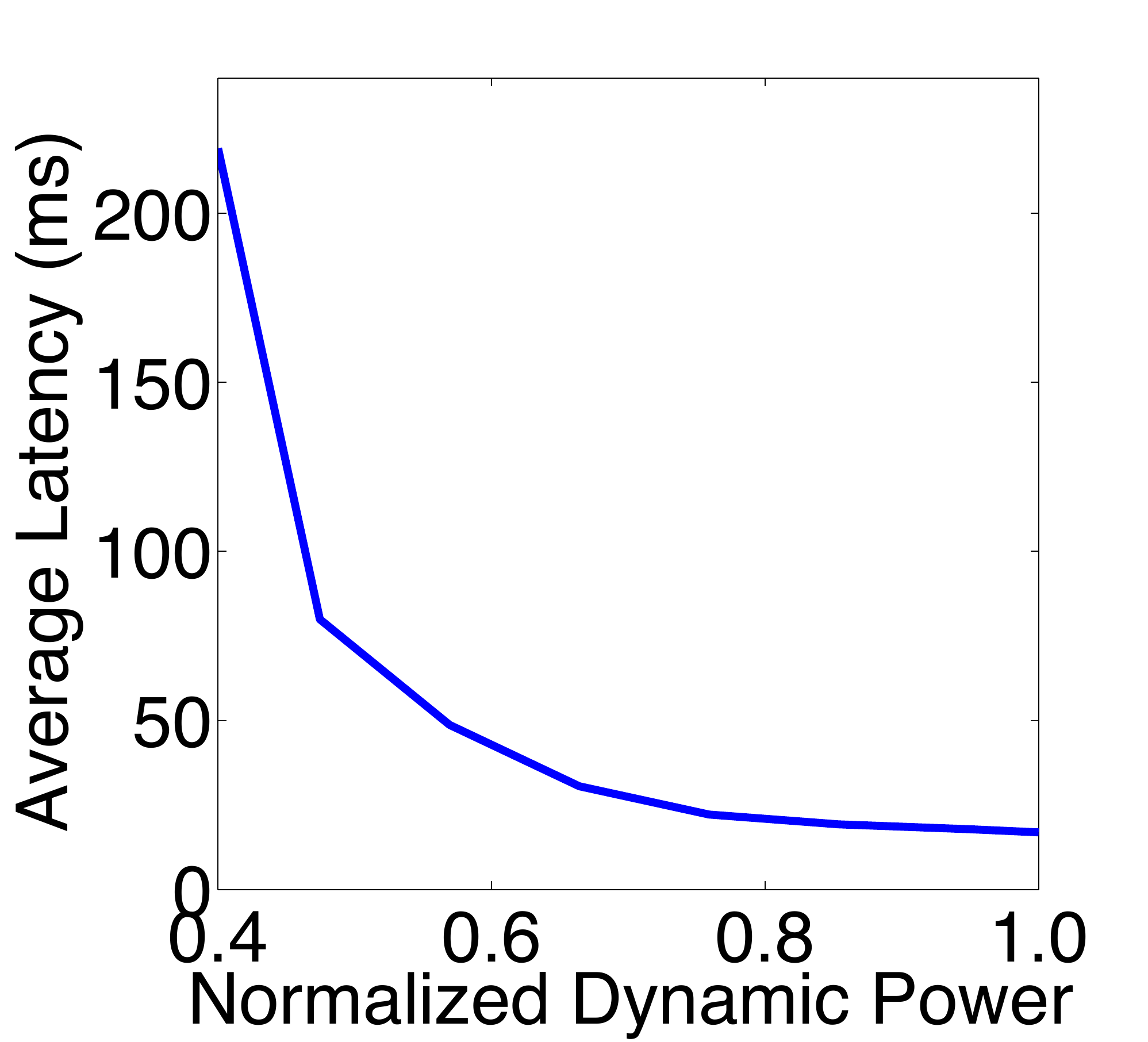}
}
\vspace{-3pt}
\protect\caption{Power utility curves for WebSearch
under 60\% load: (a) success rate, (b) average latency.}
\label{fig:Power-utility-curve}
\end{figure}

A second option is to make the power capping policy \emph{workload aware}.
A workload aware policy has knowledge of how efficiently a workload will
utilize the available power, and can predict how the intensity of the workload
will change in the near future.  As a result, a workload aware policy can
control when capping takes place such that it minimizes the 
\emph{long-term} impact on workload performance.  
As an example, we examine the
workload behavior of one of our applications, WebSearch (which models how the index searching
component of commercial web search engine services queries; see Section~\ref{sec:methodology} for details).
Figure~\ref{fig:Power-utility-curve}a shows the \emph{success rate} (i.e.,
\changesii{the percentage of requests completed within the maximum allowable service time for the workload)} 
for WebSearch as the power is reduced, while 
Figure~\ref{fig:Power-utility-curve}b shows the \emph{average latency}
for servicing queries.
In these figures, we observe that the success rate does not drop significantly, and the latency
does not increase greatly, when we start to lower the power. However, as we lower
the power further, changes in success rate and latency begin to become substantial.

A workload aware capping policy can exploit the information in
Figure~\ref{fig:Power-utility-curve} to minimize performance degradation.  For
example, suppose that we need to cap the overall dynamic power consumption of
WebSearch at 75\% of peak power over the next two execution periods.  One
approach is to cap the first period at 50\% power consumption, and then allow the
second period to execute at full power.  This has an overall success rate of
95.13\%, with an average latency of 44~ms.  A second approach is to uniformly cap
both periods at 75\% power consumption, which leads to a 99.32\% success rate and
an average latency of only 23~ms.  As a workload aware capping policy can track
the power utility of an application, it can correctly predict that the second
option is better for application performance.

A third option for power capping policy design is whether the policy should be
implemented in a \changes{\emph{centralized or decentralized}} manner.  A centralized power
capping controller resides in a single server, but controls the power of every
server within a rack.  As such, this server is aware of the workload intensity
of every server within the rack, and can use this intensity information to make
power distribution decisions.  In contrast, a decentralized controller will have
each server within the rack manage its own power.  Unlike a centralized policy,
a decentralized policy does not require communication across servers, but it
is unable to make globally-optimal decisions.  A decentralized power capping
policy is more fault tolerant than a centralized policy, which can break if the
server performing the centralized calculations fails, and is also more scalable.

\section{SizeCap: ESD Sizing Framework}

In this section, we introduce \emph{SizeCap}, an ESD sizing framework
for fuel cell powered data centers. SizeCap sizes the ESD just large
enough to cover the majority of power surges that occur within a rack, and employs power capping
techniques specifically designed for fuel cell powered data centers
to handle the remaining power surges. In this section, we first
provide an overview of SizeCap. Then, we propose five
power capping policies with various levels of system and workload
knowledge.

\subsection{Framework Overview}

\begin{figure}[t]
\centering
\includegraphics[width=0.9\linewidth]{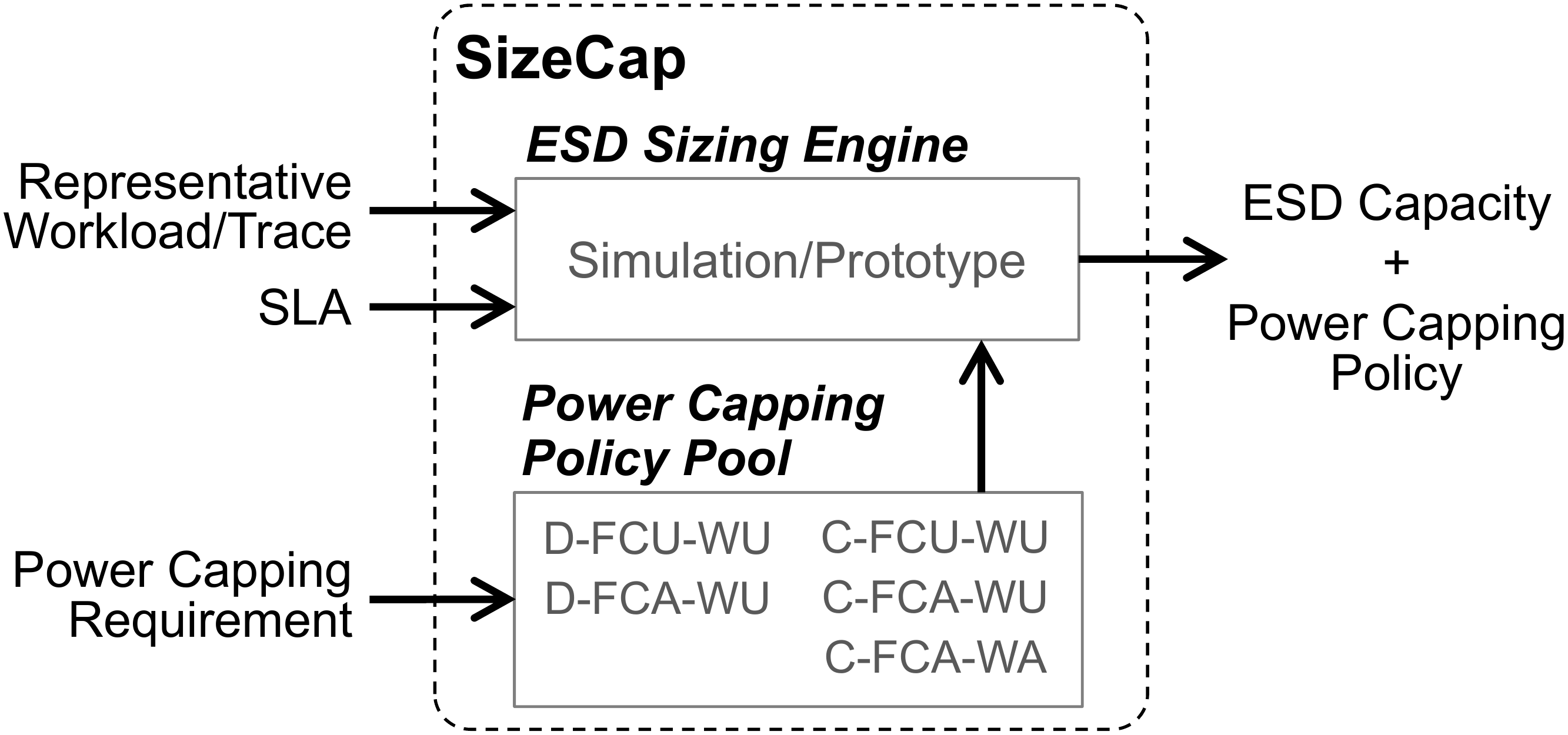}%
\protect\caption{\label{fig:SizeCap}High-level overview of SizeCap.}
\end{figure}

SizeCap uses a representative workload and trace to determine the appropriate
size of the ESD, as well as which power capping policy to use with it, as
shown in Figure~\ref{fig:SizeCap}.  SizeCap consists of two components: a
power capping policy pool, and an ESD sizing engine.

The power capping policy pool contains a list of all possible capping policies
that can be used in tandem with a smaller ESD.  Depending on the rack
configuration, some of the capping policies may not be implementable (or may not
be practical at scale). This information is relayed to the pool to
disable any unusable capping policies for this particular rack.  As we saw in
Section~\ref{sec:analysis:capping}, the capping policy can impact workload
performance, so it is important to take into account the power capping
policy that will be used when sizing the ESD.

The ESD sizing engine works to find the most underprovisioned ESD capacity
that can still satisfy workload SLAs.  The representative trace allows SizeCap to
determine the typical power surges that the data center rack will encounter.
This information, combined with the workload SLA, is used to explore various
combinations of ESD size and power capping policy, and will determine
the minimum ESD size (paired with a power capping policy) that does not violate 
any SLAs.




\subsection{Power Capping Policies}
\label{sec:sizecap:policy}
We propose five power capping policies that SizeCap can employ:
\begin{itemize}
\itemsep 0pt \parskip 0pt
\item Decentralized Fuel Cell Unaware Workload Unaware\\(D-FCU-WU)
\item Centralized Fuel Cell Unaware Workload Unaware\\(C-FCU-WU)
\item Decentralized Fuel Cell Aware Workload Unaware\\(D-FCA-WU)
\item Centralized Fuel Cell Aware Workload Unaware\\(C-FCA-WU)
\item Centralized Fuel Cell Aware Workload Aware\\(C-FCA-WA)
\end{itemize}
Note that certain combinations of the policy options from 
Section~\ref{sec:analysis:capping} are not feasible.  Workload aware policies
cannot be implemented in a decentralized manner, as each controller would need
to receive workload information from every server, incurring high communication
overhead.  Fuel cell unaware policies cannot be workload aware, as they cannot
predict when the ESD's energy will be exhausted, and are thus unable to
redistribute its usage over the duration of fuel cell ramp up.

\begin{figure}[t]
\centering
\includegraphics[width=\linewidth]{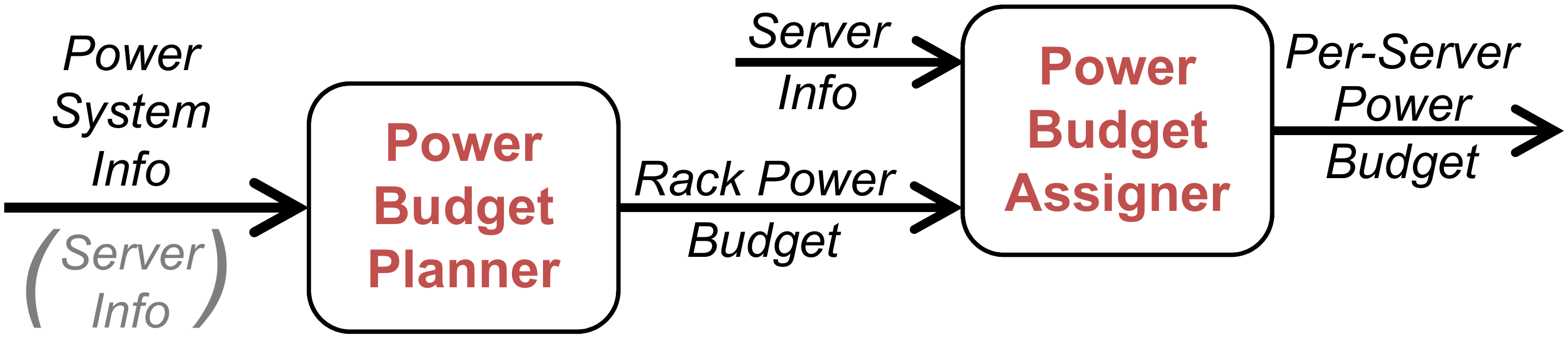}%
\caption{\label{fig:Diagram-Power-Capping}Power capping policy design.}
\end{figure}

All of these policies consist of two components: a \emph{power budget planner} and a
\emph{power budget assigner}, as shown in Figure~\ref{fig:Diagram-Power-Capping}.
At every power capping period ($T_{Capping}$), the power budget
planner determines the total rack power budget for the next period, based on
the current state of the power system (and possibly the workload information from each server in the rack).  This budget is then given to the power
budget assigner, which uses this budget along with information about each server
in the rack to determine how this budget is distributed amongst the servers in
the next period.  This information differs with each policy, as we will 
discuss later.

\begin{figure}[h]
\centering
\includegraphics[width=7.15cm]{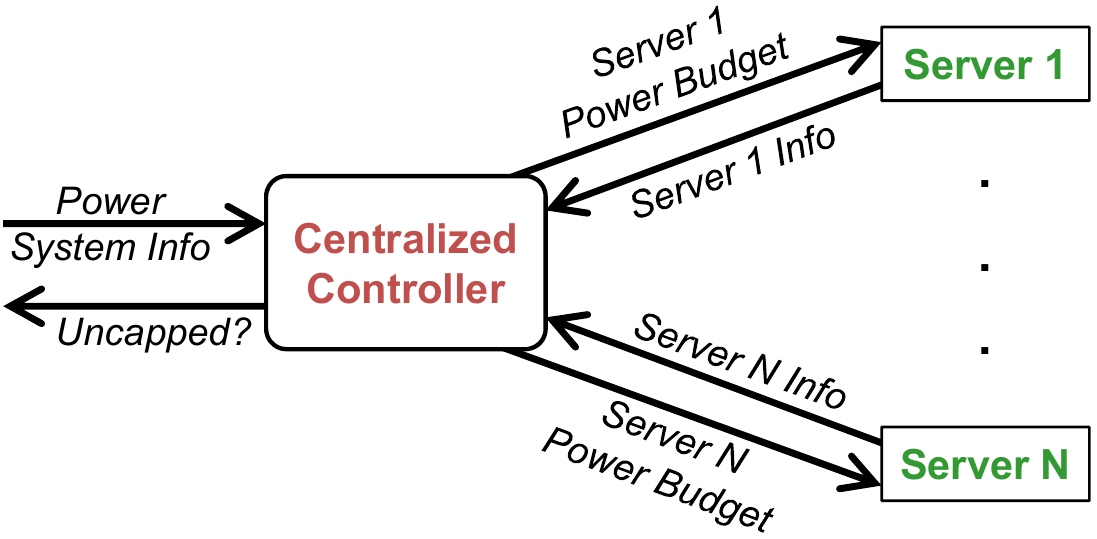}%
\vspace{2pt}
\caption{\label{fig:centralized}Centralized power capping policy design.}
\end{figure}

\begin{figure}[h]
\centering
\includegraphics[width=7.15cm]{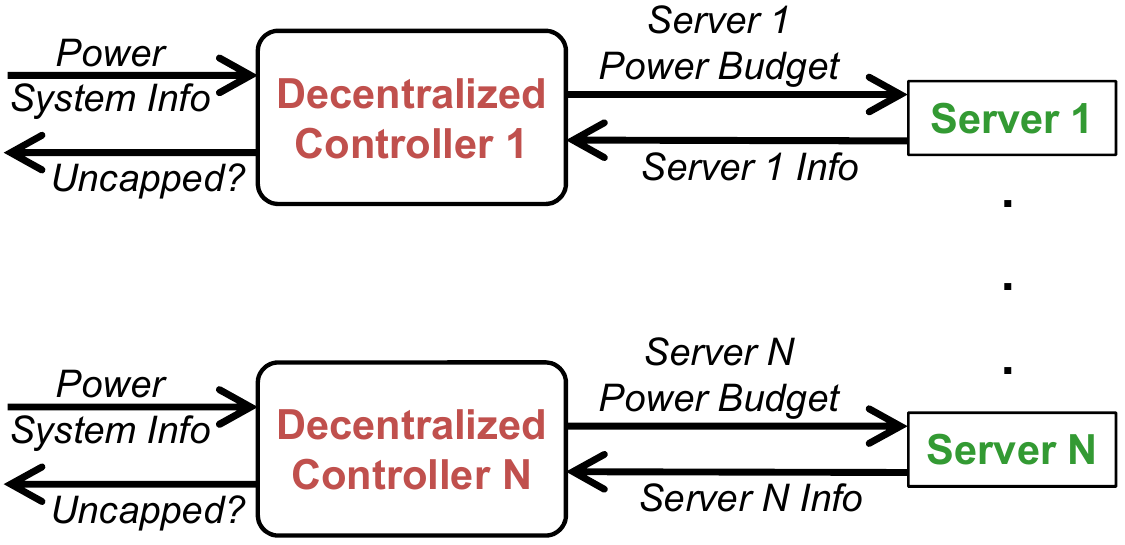}%
\vspace{2pt}
\caption{\label{fig:decentralized}Decentralized power capping policy design.}
\vspace{4pt}
\end{figure}

Each power capping policy can use either a centralized \changes{controller} or 
decentralized controllers.  A centralized controller, as shown in 
Figure~\ref{fig:centralized}, collects information from the power system and
from all of the servers, allowing it to have a global view
of the current system state.  A decentralized mechanism, as shown in
Figure~\ref{fig:decentralized}, instead assigns a separate controller to each
server in the rack.  Each controller is aware of only the power system state,
as well as the state of the server that it resides on, but is \emph{unaware} of
the state of the other servers.  A decentralized controller can only
cap the power of the one server that it is assigned to.

\vspace{6pt}

In both centralized and decentralized mechanisms, the ESD controller must know when all of the servers are not
being capped, as this informs the ESD that it can recharge itself (as the load
demand has been fully met).  In the centralized mechanism, the central controller
sends a single packet to the ESD controller to notify it that none of the servers
are capped.  In the decentralized mechanism, \emph{each} decentralized
controller must send its own packet, notifying the ESD controller that the
individual server is not being capped, and is receiving all the power that it
has demanded.


Table~\ref{tab:notation_use} lists the variables we use to describe
the power capping policies. Here, the server workload intensity
$\lambda_{i}$ characterizes how intensive the workload running on each server is, and is approximated with the request arrival
rate \emph{normalized} to the maximum request arrival rate that the server can 
handle.\changes{\footnote{\changes{The maximum request arrival rate that a server can
handle for a given workload is typically known, or can be profiled.}}}
The fuel cell state $S_{Fuel\: Cell}$
includes information about the fuel cell output power, fuel flow rate, and the
hydrogen, oxygen, and water pressure within the fuel cell stack. $P_{idle}$
characterizes the power consumption when the server has a workload
intensity of zero
\changes{(i.e., it is not servicing any requests)}.
We define $t$
to be the current time \changes{period}, and $t+k$ represents the $k$th power capping period
after the current time.
\changes{Therefore, $P_{Rack}\left(t\right)$ represents the current rack power, and 
$P_{Rack}\left(t+k\right)$ represents the rack power $k$ power capping periods 
in the future.  Other variables also follow this time period convention.}

\setlength{\extrarowheight}{1.5mm}
\begin{table}[t]
\centering
{\footnotesize
\begin{tabular}{|l|l|}
\hline 
\textbf{Notation} & \textbf{Meaning}\\
\hhline{|=|=|}
$P_{Rack}$ & Rack power \\
\hline
$\widetilde{P}_{Rack}$ & Rack power budget\\
\hline 
$P_{Server\: i}$ & Server $i$'s power\tabularnewline
\hline 
$\widetilde{P}_{Server\: i}$ & Server $i$'s power budget\tabularnewline
\hline 
$\lambda_{i}$ & Server $i$'s workload intensity\\
\hline 
$P_{idle}$ & Server idle power\\
\hline 
$N$ & Number of servers within the rack\\
\hline 
$S_{Fuel\, Cell}$ & Fuel cell state\\
\hline 
$P_{Following}$ & Fuel cell load following rate\\
\hline 
$E_{ESD}$ & Energy currently stored within the ESD\\
\hline 
$E_{min}$ & ESD energy threshold\\
\hline 
$\eta$ & ESD charging/discharging efficiency\\
\hline 
$T_{Capping}$ & Power capping period\\
\hline 
\end{tabular}
}
\vspace{5pt}
\protect\caption{\label{tab:notation_use}Variables used to define power capping policies.}
\end{table}

\subsubsection{Fuel Cell and Workload Unaware Policies}
\label{sec:sizecap:policies:unaware}

The fuel cell unaware, workload unaware
power capping policies (\mbox{C-FCU-WU} and D-FCU-WU) do \emph{not} take
advantage of the load following behavior of the fuel cell, nor do they
take workload characteristics into account. Since
these policies are workload unaware, when the load demand suddenly
increases, they do not evaluate the impact that power
provisioning \changes{has} on current and future workload performance. Instead, they
just try to ramp up the rack power as fast as possible, such that
the fuel cell can sense this fast-increasing load and also increase
its power as fast as possible, thus reducing the gap between fuel cell
power and load. However, when the ESD energy drops down
to $E_{min}$,\footnote{As stated in Section~\ref{sec:analysis}, the ESD
maintains a minimum amount of energy to ensure the normal operation of the UPS.}
these policies have no choice but to cap the rack power at the
load following rate.\footnote{The load following rate is the slope at which
a fuel cell system will never experience a power shortfall within the rack power operating range, as defined in Section~\ref{sec:analysis}.}
This guarantees that \changes{no further power is extracted from the ESD}. 

Thus, the power budget planner needs to know only
the current fuel cell power ($P_{FuelCell}\left(t\right)$), and the amount
of energy currently stored in the ESD ($E_{ESD}\left(t\right)$). As Equation~\ref{eq:5} shows, when
$E_{ESD}\left(t\right)>E_{min}$, the policies calculate
the maximum power that the ESD can deliver on top of the fuel cell output power
($P_{Fuel\: Cell}\left(t\right)$) for the next
power capping period ($T_{Capping}$). Note that we take into account
the energy efficiency of ESD discharging ($\eta$).
When $E_{ESD}\left(t\right)=E_{min}$, the policy
ramps up the rack power at the load following rate ($P_{Following}$). 

{\footnotesize
\begin{equation}\hspace{-3.5pt}
\widetilde{P}_{Rack}\left(t+1\right)=\begin{cases}
    P_{Fuel\, Cell}\left(t\right)+\eta\frac{E_{ESD}\left(t\right)-E_{min}}{T_{Capping}} & \text{if}\: E_{ESD}\left(t\right)>E_{min}\\
    P_{Fuel\: Cell}\left(t\right)+P_{Following}\times T_{Capping} & \text{if}\: E_{ESD}\left(t\right)=E_{min}
\end{cases}\raisetag{0pt}\label{eq:5}
\end{equation}
\vspace{3pt}
}

This power budget planner can
be implemented in either a centralized or decentralized manner. The
decentralized implementation runs a copy of the same power budget planner
on each server.

The power budget assigner differs between the centralized and decentralized
versions of the policy.  For the centralized policy, the assigner requires
 the current workload intensity ($\lambda_{i}\left(t\right)$) from each server,
and divides the rack power budget based on the workload
intensity of each server, as a server with a greater workload intensity usually demands
more power. The power budget assigner first assigns the
idle power ($P_{idle}$) to
each server, and then assigns the remaining dynamic power proportionally
based on each server's workload intensity (here, $N$ is the number of servers
in the rack). The power budget assigned to server $i$
during the next power capping period ($\widetilde{P}_{Server\, i}\left(t+1\right)$) is given in Equation~\ref{eq:2-1}:

\begin{footnotesize}

\begin{equation}
\widetilde{P}_{Server\, i}\left(t+1\right)=P_{idle}+\left(\widetilde{P}_{Rack}\left(t+1\right)-N\times P_{idle}\right)\times\frac{\lambda_{i}\left(t\right)}{\sum_{i=1}^{N}\lambda_{i}\left(t\right)}\label{eq:2-1}
\end{equation}
\vspace{5pt}

\end{footnotesize}

In reality, when we assign the server power budget based on Equation~\ref{eq:2-1},
the assigned power may exceed the server's uncapped load.
To tackle such situations, for each server, the power budget assigner
compares the server power budget assigned by Equation~\ref{eq:2-1}
with the actual power demanded by the server currently, and first only assigns enough
power to meet the demand. After that, if the rack power budget has not been
fully assigned, it assigns the remaining power budget proportionally
to each server that has not received its demanded power based on
its workload intensity.
When every server
receives its demanded power but the rack power budget has not been
fully assigned, 
\changes{we assign the remaining budget proportionally to each server based on its
current workload intensity.  Thus, the leftover power is used to increase the
power cap of each server, such that if the number of incoming requests to a
server increases significantly in the next capping period, the server can use
this extra power to mitigate the negative impact of these additional requests 
on the request success rate.}

For the decentralized policy's power budget assigner, since
it is unaware of the workload intensity of other servers, it is infeasible
to assign power based on workload intensity. Instead, it passes
the current rack power consumption ($P_{Rack}\left(t\right)$)
and the server power consumption ($P_{Server\: i}\left(t\right)$)
into a heuristic that assigns the non-idle power
to a server proportionally based on its current non-idle power consumption, as
follows:

\begin{footnotesize}
\vspace{-3pt}
\begin{equation}
\begin{split}
\widetilde{P}_{Server\, i}\left(t+1\right)=P_{idle} + \biggl(&\left(\widetilde{P}_{Rack}\left(t+1\right)-N\times P_{idle}\right) \biggr. \\
& \biggl. \times\frac{P_{Server\,\, i}\left(t\right)-P_{idle}}{P_{Rack}\left(t\right)-N\times P_{idle}} \biggr)
\end{split}
\end{equation}

\end{footnotesize}

Since each decentralized power budget planner computes the same rack
power budget $\widetilde{P}_{Rack}\left(t+1\right)$, and because

\begin{footnotesize}
\begin{equation}
P_{Rack}\left(t\right)=\sum_{i=1}^{N}P_{Server\: i}\left(t\right)
\end{equation}
\end{footnotesize}

\noindent
it can be proven that with this heuristic-based approach, the sum of each server's power budget
$\widetilde{P}_{Server\: i}\left(t+1\right)$ is equal to the rack
power budget $\widetilde{P}_{Rack}\left(t+1\right)$. 
\changes{In case the servers crash for reasons other than power availability (e.g.,
workload consolidation, software errors)}, we can update the value of $N$
in the decentralized controllers of the remaining servers, allowing power
capping to continue working normally.

\vspace{3pt}
{\bf \noindent Scalability Analysis.} The computation time and message count sent for the centralized policy
scale linearly with the number of machines per rack. The centralized policy has a single capping
controller computing the power budgeted for each machine, and must optimize
assigned power on a per-machine basis.  The capping controller must receive
messages from each server. For the decentralized policy, the computation time does not depend on the number of 
machines, while message count scales linearly. The decentralized policy has a 
capping controller per machine, allowing parallel computation for each 
machine's power budget.  Each capping controller must exchange messages with 
the ESD controller, leading to linear scaling of message count.

\subsubsection{Fuel Cell Aware, Workload Unaware Policies}

The fuel cell aware, workload unaware
power capping policies (C-FCA-WU and D-FCA-WU) improve over the 
previous two policies by adding knowledge of the load following behavior
of fuel cells into the policy.
As they are still workload unaware, they continue to ramp up the rack power as fast as
possible. However, since these policies are aware of the fuel cell
load following behavior, they know how the fuel cell power varies
for a given load, based on the current state of the fuel cell.  Moreover,
these policies can calculate the change in stored ESD energy.
Therefore, they can ramp the rack power up at the  fastest potential speed,
without violating the ESD energy threshold constraint.

The power budget planner needs to know the current fuel cell state
($S_{Fuel\: Cell}\left(t\right)$) and current stored ESD energy ($E_{ESD}\left(t\right)$),
which it uses to determine the rack power budget ($\widetilde{P}_{Rack}\left(t+1\right)$) by solving
the optimization problem shown in Equation~\ref{eq:3-1}. In this
optimization problem, the power budget planner tries to maximize the
rack power budget in the next step ($\widetilde{P}_{Rack}\left(t+1\right)$).
To do this, it employs the ESD energy model ($E_{ESD\: Model}$, derived from
the fuel cell load following model; see details in Section~\ref{sec:methodology}), 
and evaluates whether ESD energy $E_{ESD}\left(t+1\right)$ at the next step
would fall below the energy threshold $E_{min}$ under a rack power budget
$\widetilde{P}_{Rack}\left(t+1\right)$.
This power budget planner can be implemented in both a centralized and
decentralized manner. A decentralized implementation needs to run a copy of the power
budget planner on each server.

\begin{footnotesize}
\vspace{-10pt}

\begin{equation}
\hspace{-3.5pt}
\begin{array}{cc}
\text{maximize}\;\widetilde{P}_{Rack}\left(t+1\right)\\
s.t.\; E_{ESD}\left(t+1\right)=E_{ESD\: Model}\left(\widetilde{P}_{Rack}\left(t+1\right),\; S_{Fuel\, Cell}\left(t\right),\,\, E_{ESD}\left(t\right)\right)\\
E_{ESD}\left(t+1\right)\geq E_{min}
\end{array}\raisetag{0pt}\label{eq:3-1}
\end{equation}

\end{footnotesize}

The power budget assigner for the fuel cell
aware, workload unaware policies is exactly the same as that of the
fuel cell unaware, workload unaware policies presented in 
Section~\ref{sec:sizecap:policies:unaware}. 
The scalability of the centralized/decentralized policies is also similar to their counterparts.

\subsubsection{Fuel Cell Aware, Workload Aware Policy}

Unlike the other four policies, the fuel cell aware, workload aware policy (C-FCA-WA)
is aware of how the provisioned power impacts future
workload performance, and tries to optimize workload performance when it makes
power capping decisions. 
\changes{This policy may
not necessarily provision the full power demanded by the rack even when the ESD energy can be utilized. 
Instead, this policy tries to intelligently distribute the ESD energy across \emph{several} periods in the near future, 
by using an estimate of future workload intensity to determine when this ESD energy is 
best spent (and thus when more aggressive capping is needed), optimizing for workload performance.
}

In order to determine how the provisioned power impacts future workload performance, the power budget planner of C-FCA-WA
relies on the workload power utility function ($f_{utility}$), which characterizes the workload
performance under different server power budgets and workload intensities.\footnote{In our implementation, the workload power utility function uses the workload request success rate as an example performance metric, 
representing the success rate as a function dependent on the server power budget under different workload intensities.}
The power budget planner collects the future workload intensity estimates for all servers within the rack ($\left\{ \lambda_{i}\left(t+k\right),\; k=1,...,P\right\}$).
Based on this information, for each power capping period in the near future,
the power budget planner uses the average server power budget 
($\frac{1}{N}\widetilde{P}_{Rack}\left(t+k\right)$)
and the average workload intensity within the rack ($\frac{1}{N}\sum_{i=1}^{N}\lambda_{i}\left(t+k\right)$)
 to approximate the average workload performance $f_{utility}$ within the rack.
After that, the power budget planner sums up the approximated
workload performance for the next $P$ power capping
steps, and tries to optimize it, which is shown as the objective
function of the following optimization problem:

\begin{footnotesize}
\vspace{-5pt}
\begin{equation}
\begin{array}{c}
\text{maximize}\sum_{k=1}^{P}f_{utility}\left(\frac{1}{N}\widetilde{P}_{Rack}\left(t+k\right),\;\frac{1}{N}\sum_{i=1}^{N}\lambda_{i}\left(t+k\right)\right)\\
s.t.\; E_{ESD}\left(t+k\right)=E_{ESD\: Model}\Big(\left\{ \widetilde{P}_{Rack}\left(t+j\right),\: j=1,...,k\right\} ,\: \\
S_{Fuel\, Cell}\left(t\right),\; E_{ESD}\left(t\right)\Big)\\
E_{ESD}\left(t+k\right)\geq E_{min}\\
k=1,...,P
\end{array}\label{eq:1}
\vspace{-2pt}
\end{equation}

\end{footnotesize}
\vspace{5pt}

\changes{
In addition, similar to the fuel cell aware, workload unaware policies, the power budget planner of C-FCA-WA must also
maintain the ESD energy threshold constraint.
To do this, it needs to know the current ESD
energy ($E_{ESD}\left(t\right)$) and fuel cell state ($S_{Fuel\: Cell}\left(t\right)$),
and leverages the ESD energy model ($E_{ESD\: Model}$, see details in Section~\ref{sec:methodology})
to ensure that the ESD energy in the near future ($\left\{ E_{ESD}\left(t+k\right),\: k=1,...,P\right\}$)
is always greater than the ESD energy threshold $E_{min}$.
}

\changes{
The power budget assigner is similar to that of
the centralized workload unaware policies. The only difference
is that since this policy is aware of future workload intensity,
instead of using the current workload intensity $\lambda_{i}\left(t\right)$
to distribute the power, it uses the workload intensity in the next
power capping period $\lambda_{i}\left(t+1\right)$ to assign the rack
power budget, as this better reflects the demanded power in the next
power capping period:
}

\begin{footnotesize}
\vspace{-11pt}

\begin{equation}
\widetilde{P}_{Server\, i}\left(t+1\right)=P_{idle}+\left(\widetilde{P}_{Rack}\left(t+1\right)-N\times P_{idle}\right)\times\frac{\lambda_{i}\left(t+1\right)}{\sum_{i=1}^{N}\lambda_{i}\left(t+1\right)}\label{eq:2}
\end{equation}
\vspace{2pt}

\end{footnotesize}

\changes{
As we can see, the fuel cell aware, workload aware policy must 
collect the workload intensity from each server to determine how its power capping decisions impact future workload performance.
Therefore, we cannot implement a decentralized version of this policy, as it would require every server to communicate its
workload performance information to every other server, which would induce a high communication overhead.
}

\vspace{3pt}
{\bf \noindent Scalability Analysis.} \changes{Similar to both C-FCU-WU and \mbox{C-FCA-WU}, the computation time and message count of this policy scale} linearly with the number of machines per rack.


\section{Evaluation Methodology}
\label{sec:methodology}


{\noindent\bf System Configuration.} We study one rack of servers, powered by a single
12.5~kW fuel cell system with a load following rate
of 16~W/s. The rack consists of 45 identical
dual socket production servers with 2.4GHz Intel Xeon CPUs. Each
server runs a power capping software driver developed in-house, leveraging
Intel processor power management capabilities. 
In every power capping period ($T_{Capping}=2s$), the 
power capping controller notifies the driver on each server to set its power budget.
Information about the current fuel cell state, ESD energy, and rack power are measured, and each server
can poll them via an Ethernet interface.
The ESD in our study has a 95\% charging/discharging efficiency~\cite{burke2007batteries}. 


\vspace{3pt}
{\noindent\bf Workload and Traces.} We use both synthesized single power surge traces and production data center workload traces collected from Microsoft data centers,\footnote{The production data center workload traces are processed and scaled to fit our configuration.} which capture real workload intensity and power profiles.
We use WebSearch for our production workload. WebSearch is an internally
developed workload to emulate the index searching component of commercial
search engines~\cite{luo2014characterizing}. It faithfully accounts for queuing, delay variation, and request dropping.


\vspace{3pt}
{\noindent\bf Metrics.} We use the \emph{success rate} and \emph{average latency} of search requests to represent the overall workload performance
for both synthesized traces and production data center traces. 
\changesii{Success rate characterizes the percentage of requests completed within the maximum allowable service time for the workload.} 
Average latency
characterizes the average service latency of all requests. In addition, we also evaluate 95\textsuperscript{th} percentile (P95)
latency, as tail latency is very important for search workloads.

\begin{figure*}[b]
\vspace{3pt}
\centering
\begin{minipage}{0.305\linewidth}%
\centering
\includegraphics[width=0.92\linewidth]{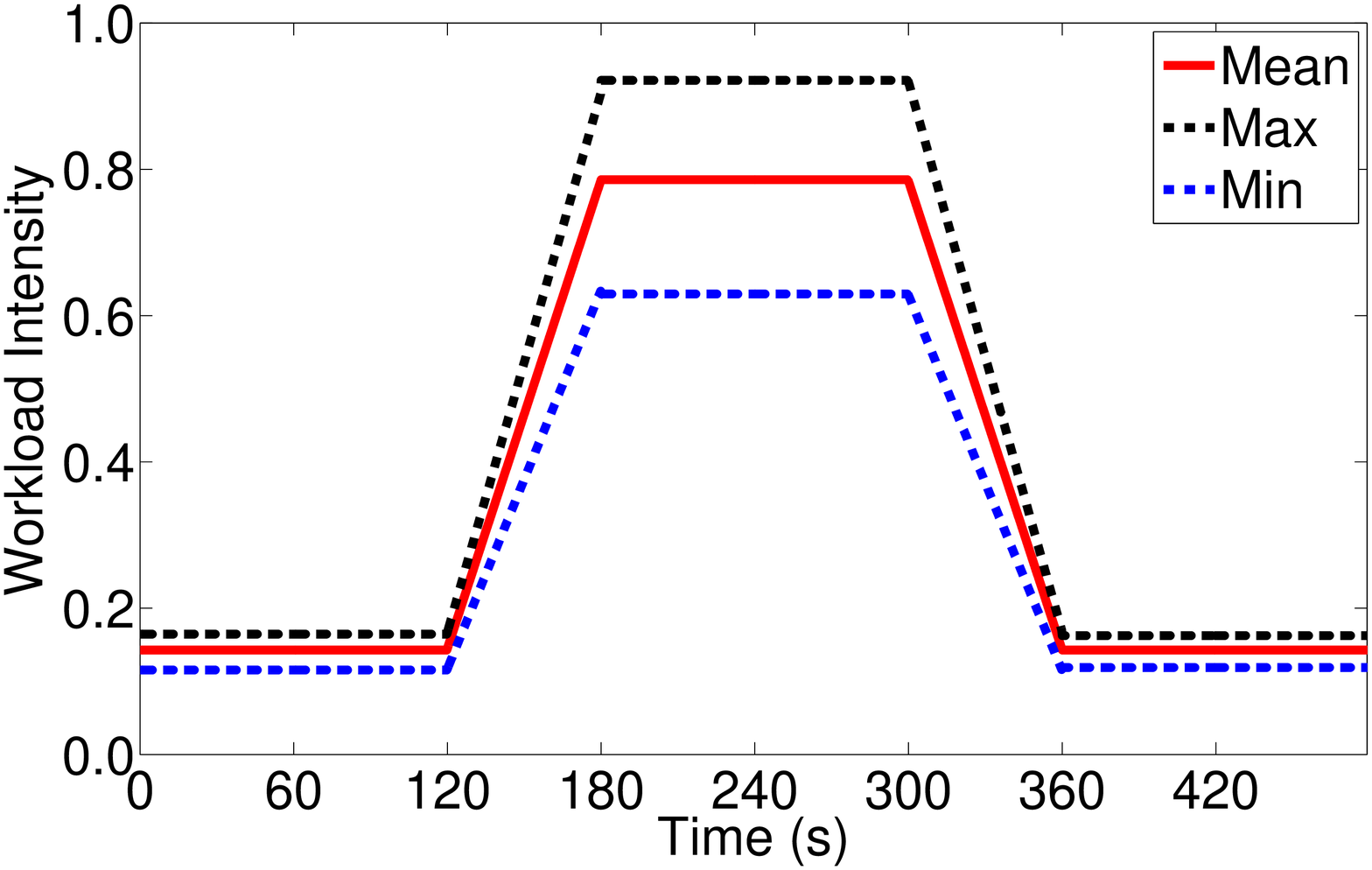}%
\vspace{-2.7pt}
\protect\caption{\label{fig:A-typical-load}Load surge model used to synthesize our evaluation trace.}
\vspace{1.093em}
\end{minipage}%
\qquad
\begin{minipage}{0.305\linewidth}%
\centering
\vspace{-3pt}
\includegraphics[width=\linewidth]{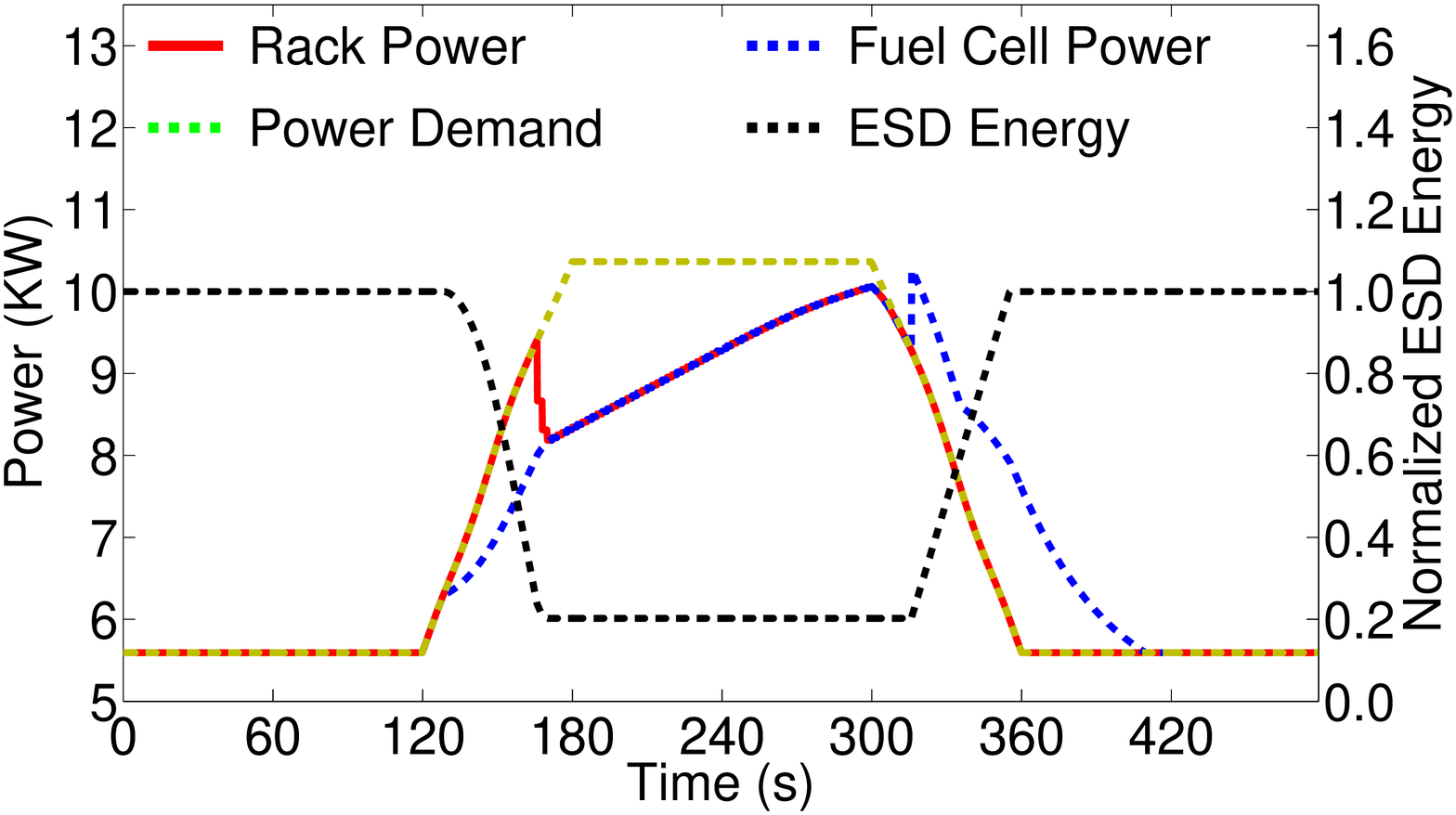}%
\vspace{-2pt}
\protect\caption{Synthetic trace behavior of SizeCap with D-FCU-WU
policy, using a 50\% underprovisioned ESD.}
\label{fig:behavior-D-FCU-WU}
\end{minipage}%
\qquad
\begin{minipage}{0.305\linewidth}%
\centering
\vspace{-1pt}
\includegraphics[width=\linewidth]{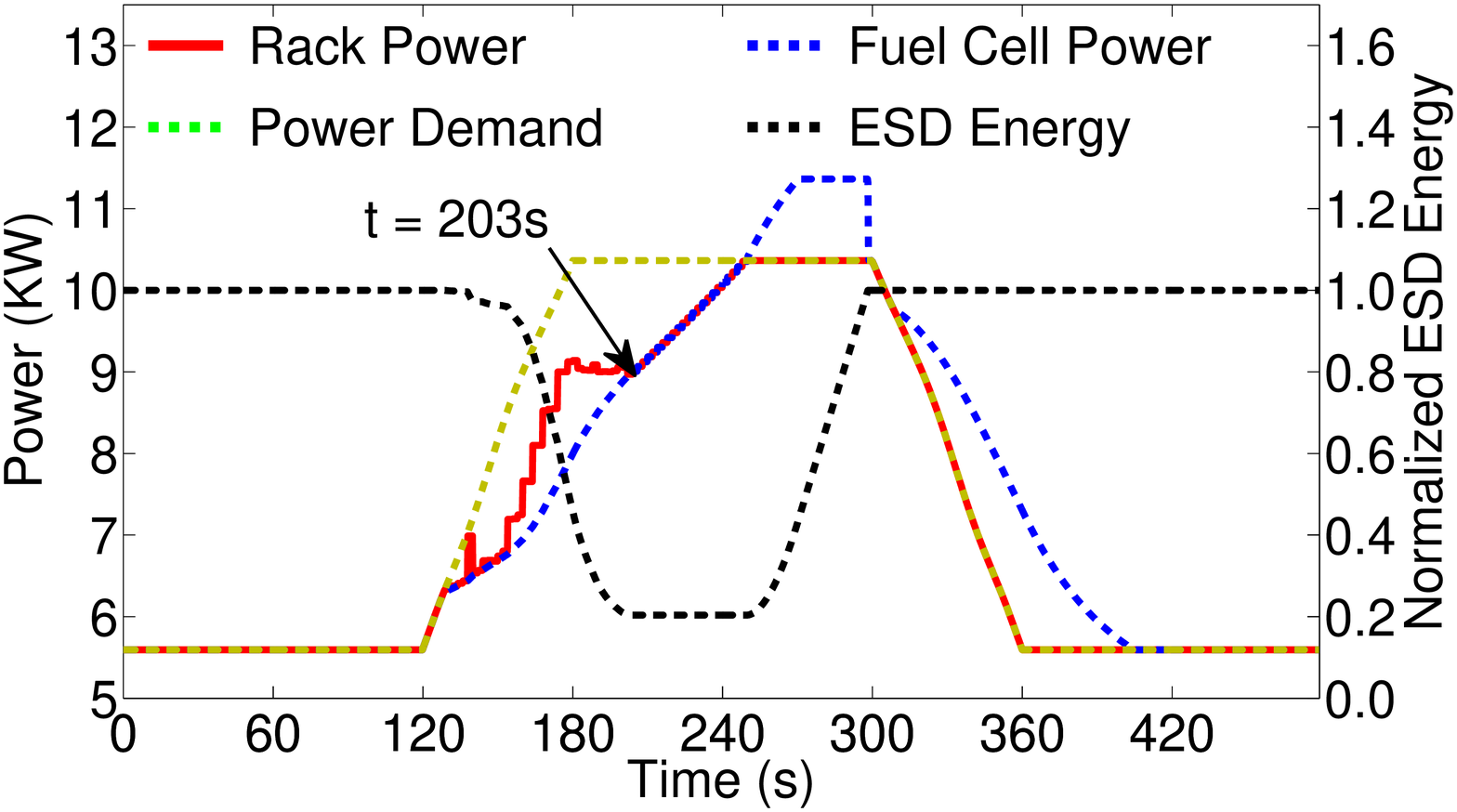}%
\vspace{-4pt}
\protect\caption{Synthetic trace behavior of SizeCap with C-FCA-WA
policy, using a 50\% underprovisioned ESD.}
\label{fig:behavior-C-FCA-WA}
\end{minipage}%
\vspace{-3pt}
\end{figure*}


\vspace{3pt}
{\noindent\bf Simulation Methodology.} We faithfully model the data center configuration described in Section~\ref{sec:background:datacenter}. 
Our simulator consists of three modules: a power capping module, a workload module, and a power system
module. The power capping module models the behavior of each power
capping policy. The workload module models the overall workload performance
and rack power consumption under power capping decisions. To do this,
we profile WebSearch on our production servers, and build a lookup
table for workload performance and server power consumption under
different workload intensities and server power budgets. This lookup
table allows the module to use the workload performance and
power consumption for each server to calculate \emph{overall} workload
performance and \emph{rack} power consumption. The power system module models
both fuel cell and ESD behavior. The \changesii{\emph{fuel cell model}}
is based on previously published fuel cell models~\cite{padulles2000integrated,Fuel_Cell_Controller_Model,mufford1999power}, and
models every component of the fuel cell system in detail (see appendix for details). Using this model, the power system module can determine
how the fuel cell state (e.g., fuel cell power, fuel flow rate, hydrogen/oxygen/water
pressure within the fuel cell stack) evolves with a given rack power. 
\changesii{The \emph{ESD energy model} uses the
fuel cell power and rack power from the fuel cell model, along with the
ESD characteristics (i.e., charge/discharge efficiency and energy capacity), to determine how much power
needs to be charged/discharged from the ESD, and how the ESD energy changes over time.
We employ the ESD energy model in our fuel cell aware power capping policies in Section~\ref{sec:sizecap:policy}.} 
The power system module assumes that the ESD energy can be measured with a precision of only 1\%
of the ESD energy capacity, to reflect the practical limitations of ESD energy measurement.  The \emph{measured} ESD energy is derived by
rounding down the \emph{actual} ESD energy to match the available precision.
The power system module 
feeds all of this information back to the power capping module for power capping decisions.


\begin{figure*}[t]
\begin{minipage}{0.305\linewidth}%
\centering
\includegraphics[width=\linewidth]{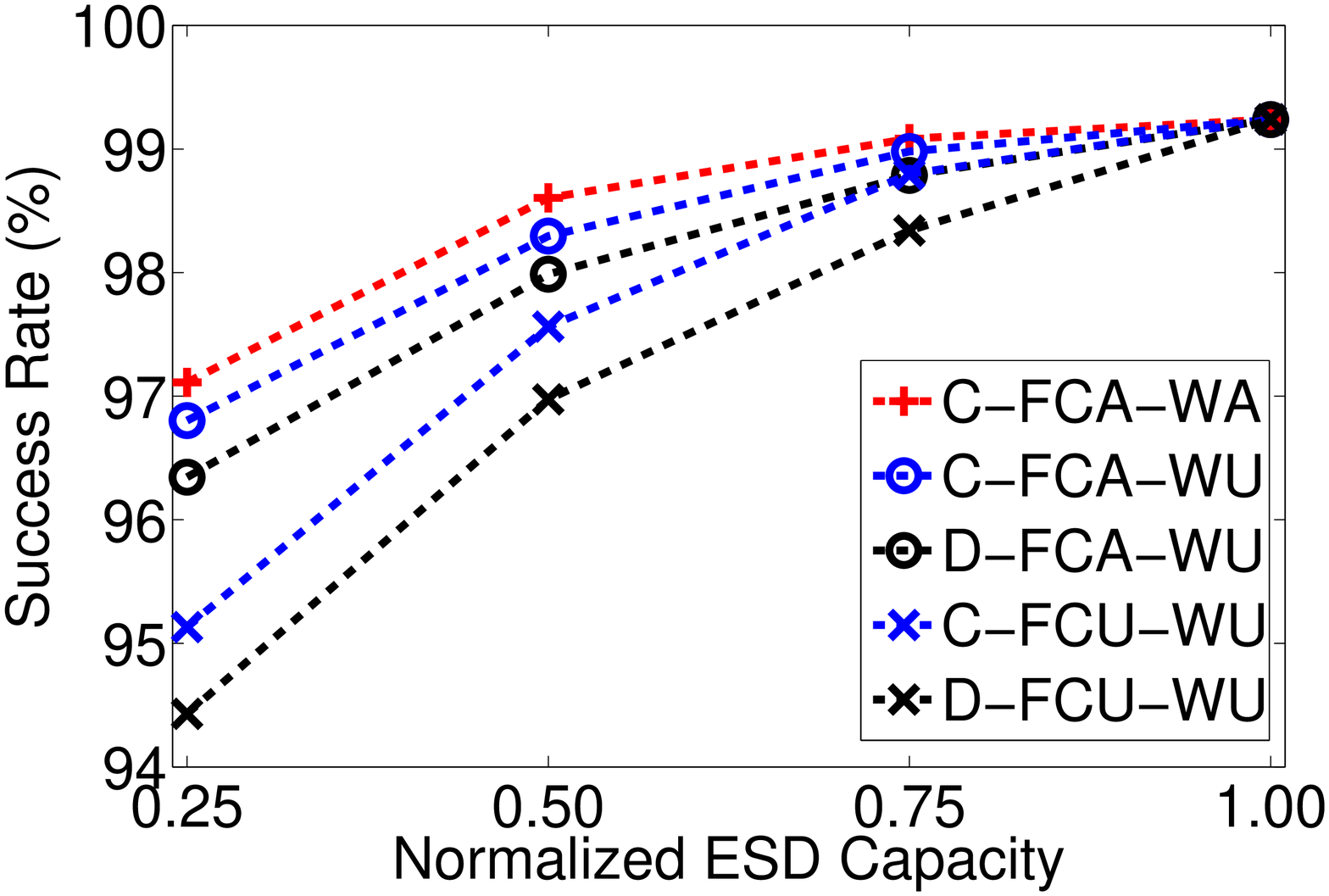}%
\vspace{-4pt}
\caption{\label{fig:success-rate-capacity}Success rate vs.\ ESD capacity for capping policies.}
\end{minipage}%
\qquad
\begin{minipage}{0.305\linewidth}%
\centering
\hspace{2pt}
\includegraphics[width=0.96\linewidth]{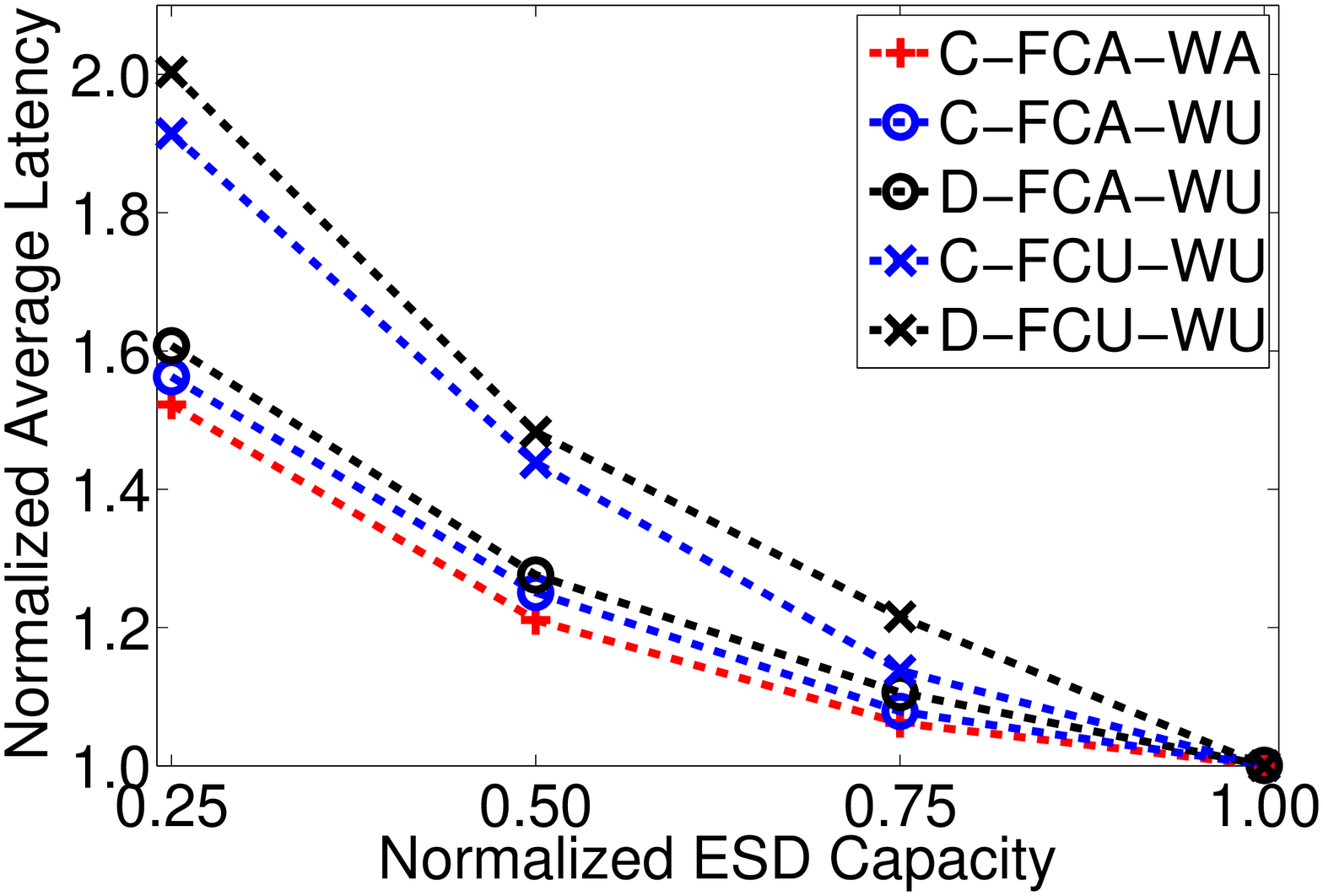}%
\vspace{-2.5pt}
\caption{\label{fig:latency-capacity}Average latency vs.\ ESD capacity for capping policies.}
\end{minipage}%
\qquad
\begin{minipage}{0.305\linewidth}%
\centering
\includegraphics[width=0.95\linewidth]{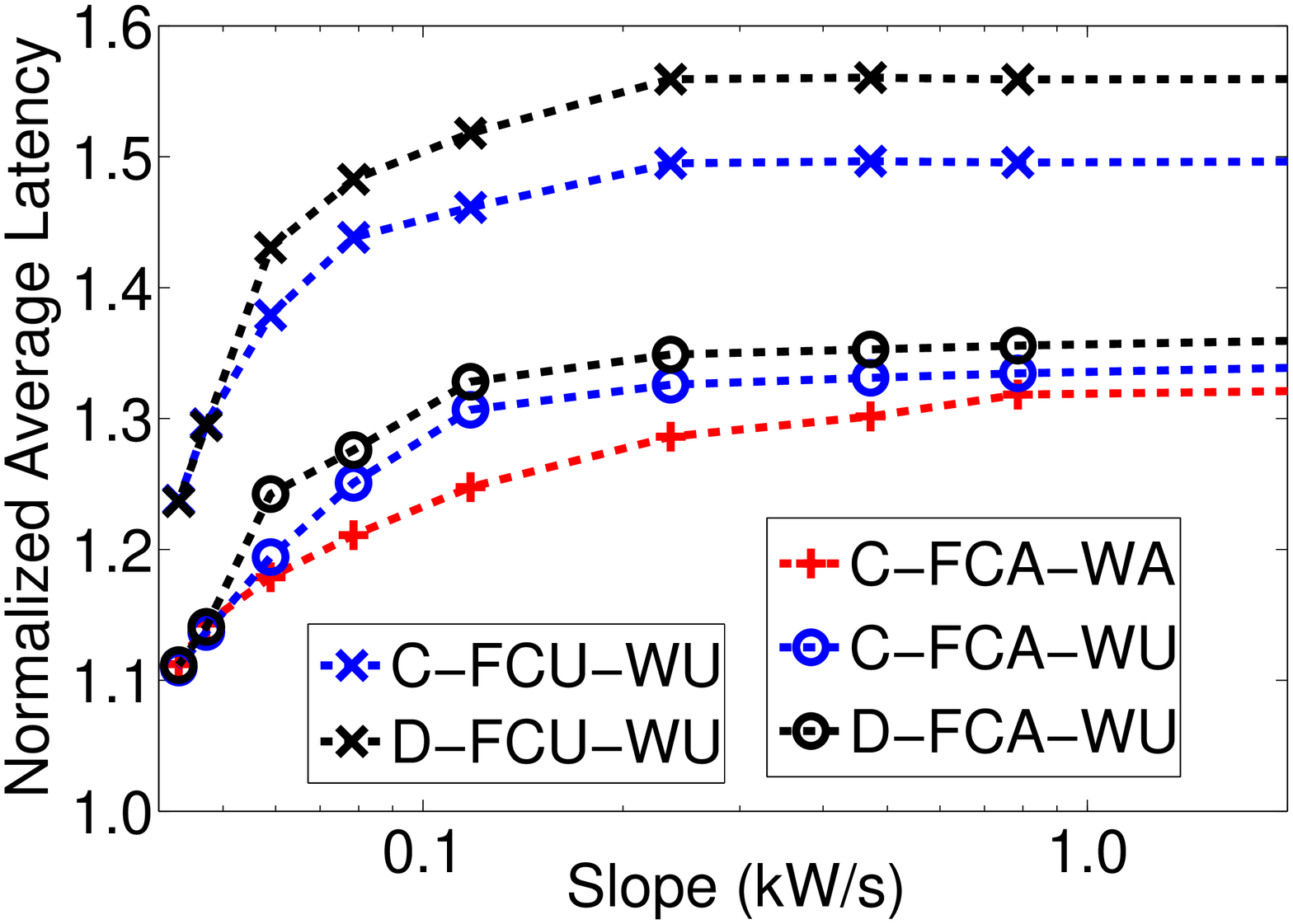}%
\vspace{-3pt}
\caption{\label{fig:latency-slope}Average latency vs.\ surge slope for capping policies.}
\end{minipage}%
\end{figure*}

\section{Experimental Results}

In this section, we evaluate the performance of SizeCap, studying how the request
success rate and average request latency change as a result of our different power
capping policies when ESD capacity is underprovisioned. We study the
impact of SizeCap first on synthetic traces, and then on traces
from production data center workloads.


\subsection{Synthesized Single Power Surge Trace}

We first use a synthesized trace, containing a single power surge, to explore 
how the combination of ESD sizing and our power capping policies impacts data
center behavior and availability.  In order to simulate workload intensity
variation within a rack, we construct our trace using the surge bounds shown in
Figure~\ref{fig:A-typical-load}.  We use insight from prior work on the workload
intensity heterogeneity across servers as measured in a production data 
center~\cite{huang2014characterizing}, and generate a trace using a normal
distribution with 8\% standard deviation from the mean surge.  The workload
intensity of each server is updated every three minutes, redistributing the
heterogeneity.

We first focus on the two extremes for our proposed policies: D-FCU-WU and 
C-FCA-WA.  Figure~\ref{fig:behavior-D-FCU-WU} shows the power demand and
rack power output for D-FCU-WU, using an ESD underprovisioned at 50\% (i.e.,
it only has enough charge to tolerate a power shortfall half the size of the 
worst-case shortfall).  Initially, the D-FCU-WU policy allows the delivered rack
power to equal the demanded power, as it discharges the ESD.  At t=180s, when
the ESD cannot provide any more power, the policy reduces the rack power by 
2~kW.  At this point, D-FCU-WU can only ramp up the output power at an average
rate of 15~W/s,
\changes{leading to significant performance degradation.  We conclude that 
D-FCU-WU is a poor policy for servicing this trace.}

Figure~\ref{fig:behavior-C-FCA-WA} shows the power demand and rack power output
for C-FCA-WA, again with a 50\% underprovisioned ESD.  Initially, C-FCA-WA does 
not follow the demanded load as closely as D-FCU-WU did, but it can gradually 
increase rack power output without ever dropping it like D-FCU-WU did,
which benefits workload performance.
\changesii{Compared with \mbox{D-FCU-WU}, C-FCA-WA has three performance advantages. First, as
C-FCA-WA is workload aware, it} knows that the workload performance drops superlinearly
with power reduction at higher loads, and so it tries to balance power capping
throughout the duration of the ramp up.  
\changesii{Second, as \mbox{C-FCA-WA}}
is aware of the load following behavior of the fuel cell, it can ramp up
the rack power budget as fast as possible when the ESD is unable provide any more power (starting at
t = 203s). In this phase, C-FCA-WA can ramp up the
output power at an average rate of 30~W/s, which is much faster than D-FCU-WU and benefits workload performance.
\changesii{Third,}
as C-FCA-WA is centralized,
it has knowledge of the actual current workload intensity of each 
server, allowing it to make global decisions, while the decentralized D-FCU-WU 
may waste power since, without the global knowledge, a server may be assigned
more power than it actually requires.

\begin{figure*}[t]
\vspace{-4pt}
\centering
\begin{minipage}{0.305\linewidth}
\centering
\includegraphics[width=\linewidth]{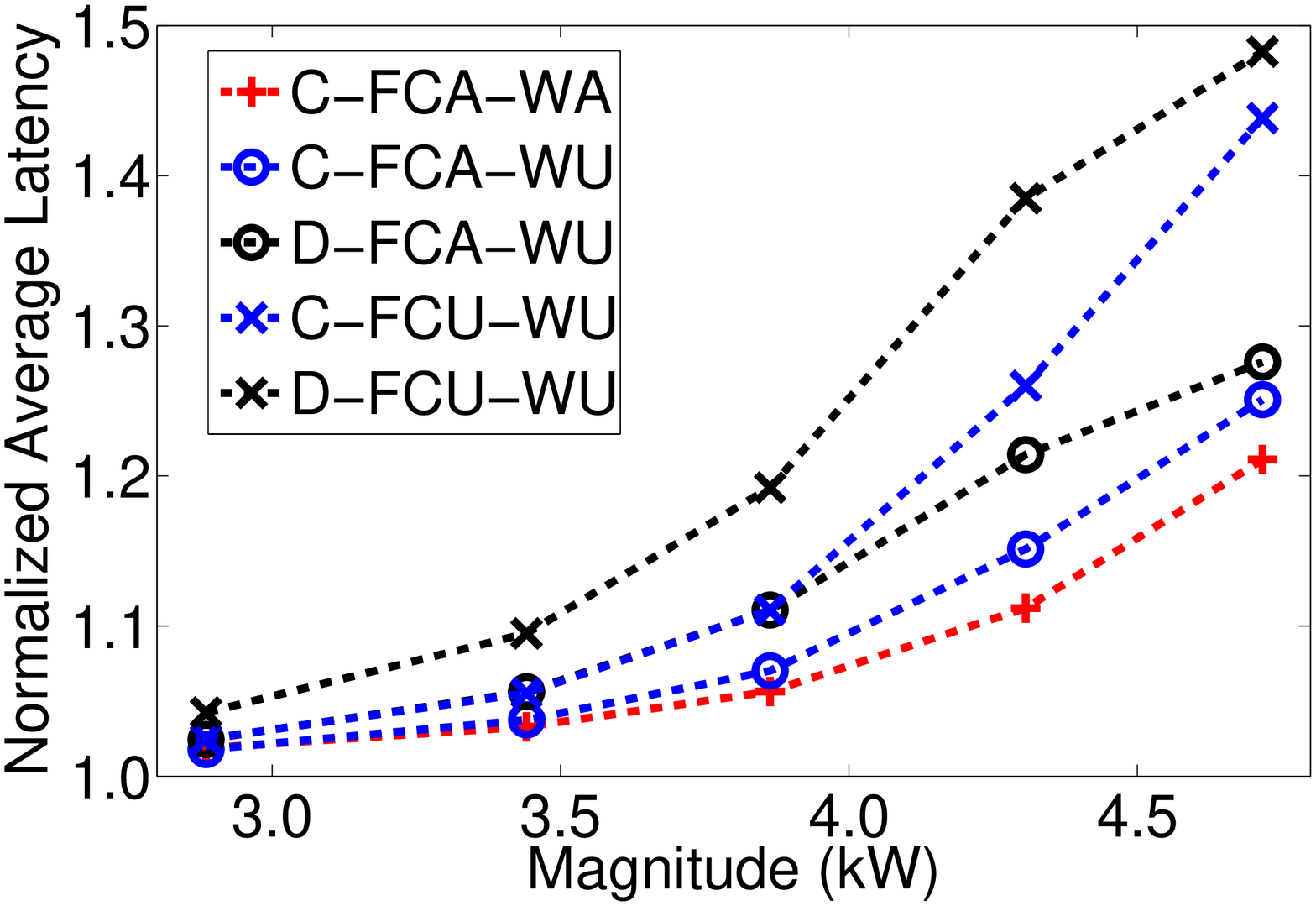}%
\vspace{-3pt}
\caption{\label{fig:latency-magnitude}Average latency vs.\ surge magnitude for capping policies.}
\end{minipage}%
\qquad
\begin{minipage}{0.305\linewidth}%
\centering
\includegraphics[width=0.95\linewidth]{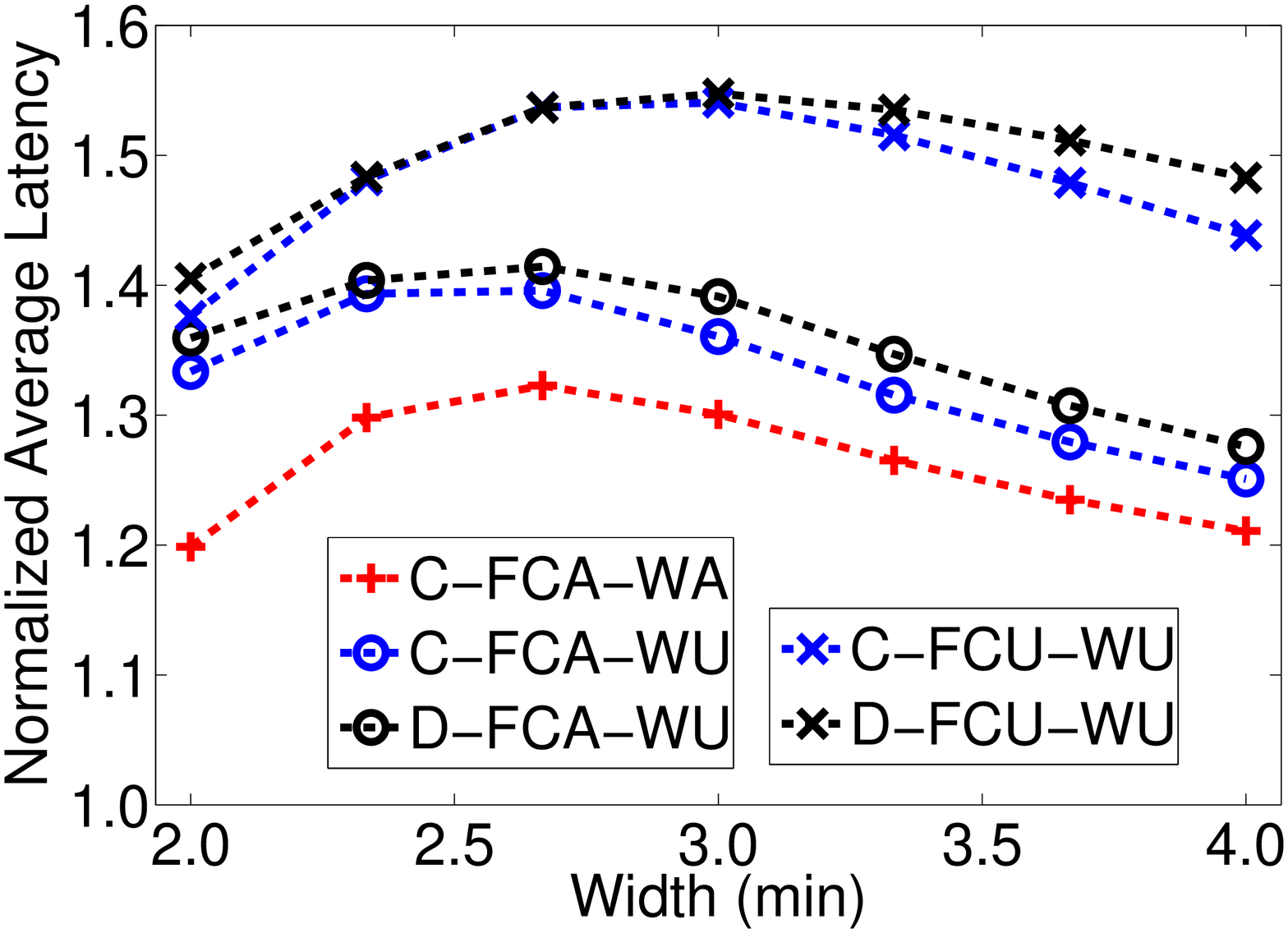}%
\vspace{-2pt}
\caption{\label{fig:latency-width}Average latency vs.\ surge width for capping policies.}
\end{minipage}%
\qquad
\begin{minipage}{0.305\linewidth}%
\centering
\includegraphics[width=0.97\linewidth]{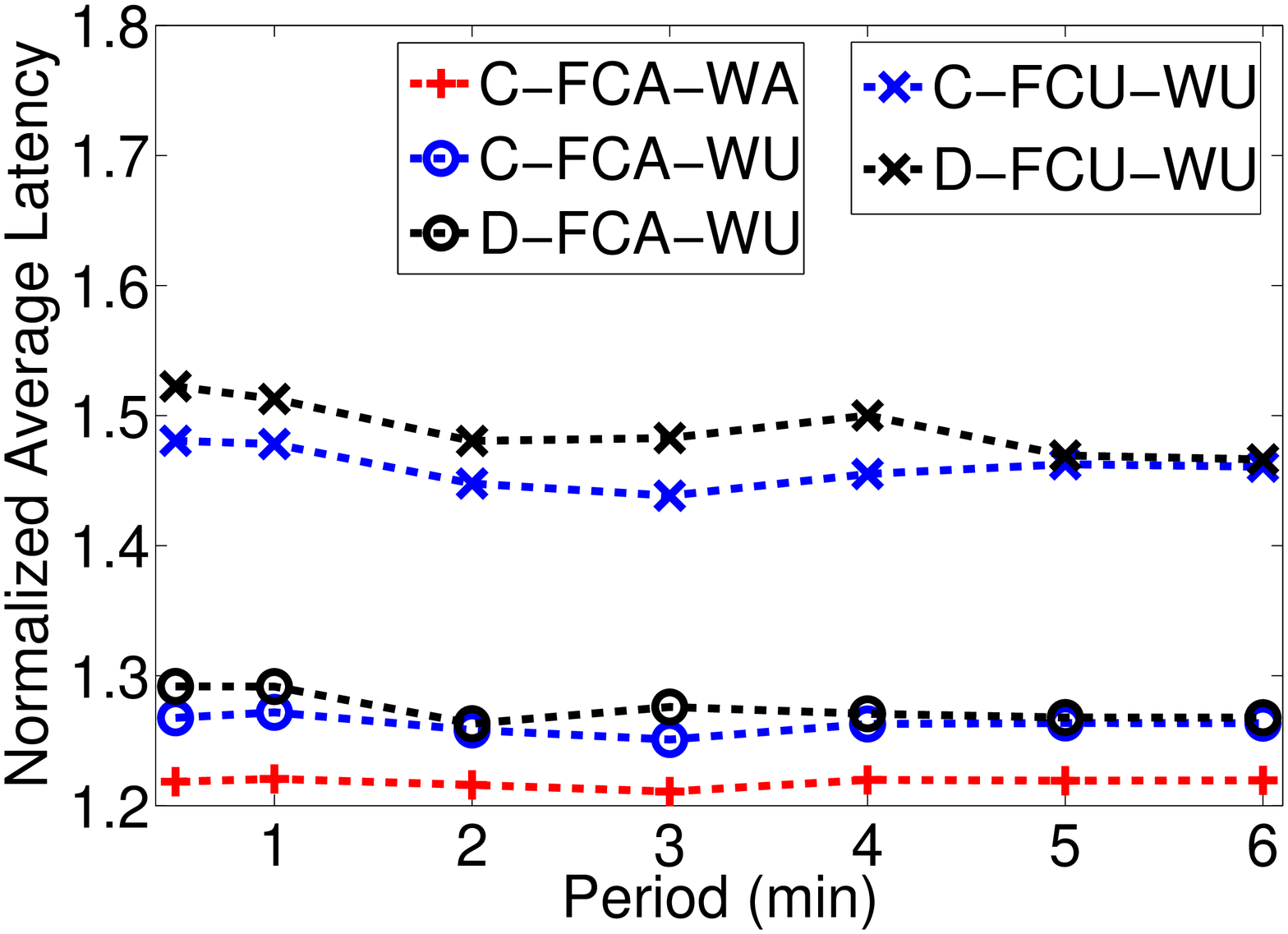}%
\vspace{-3.5pt}
\caption{\label{fig:latency-period}Latency vs.\ heterogeneity update frequency for capping policies.}
\end{minipage}%
\end{figure*}

\begin{figure*}[t]
\vspace{-7pt}
\subfloat[Success rate.]{\centering
\includegraphics[width=5.6cm]{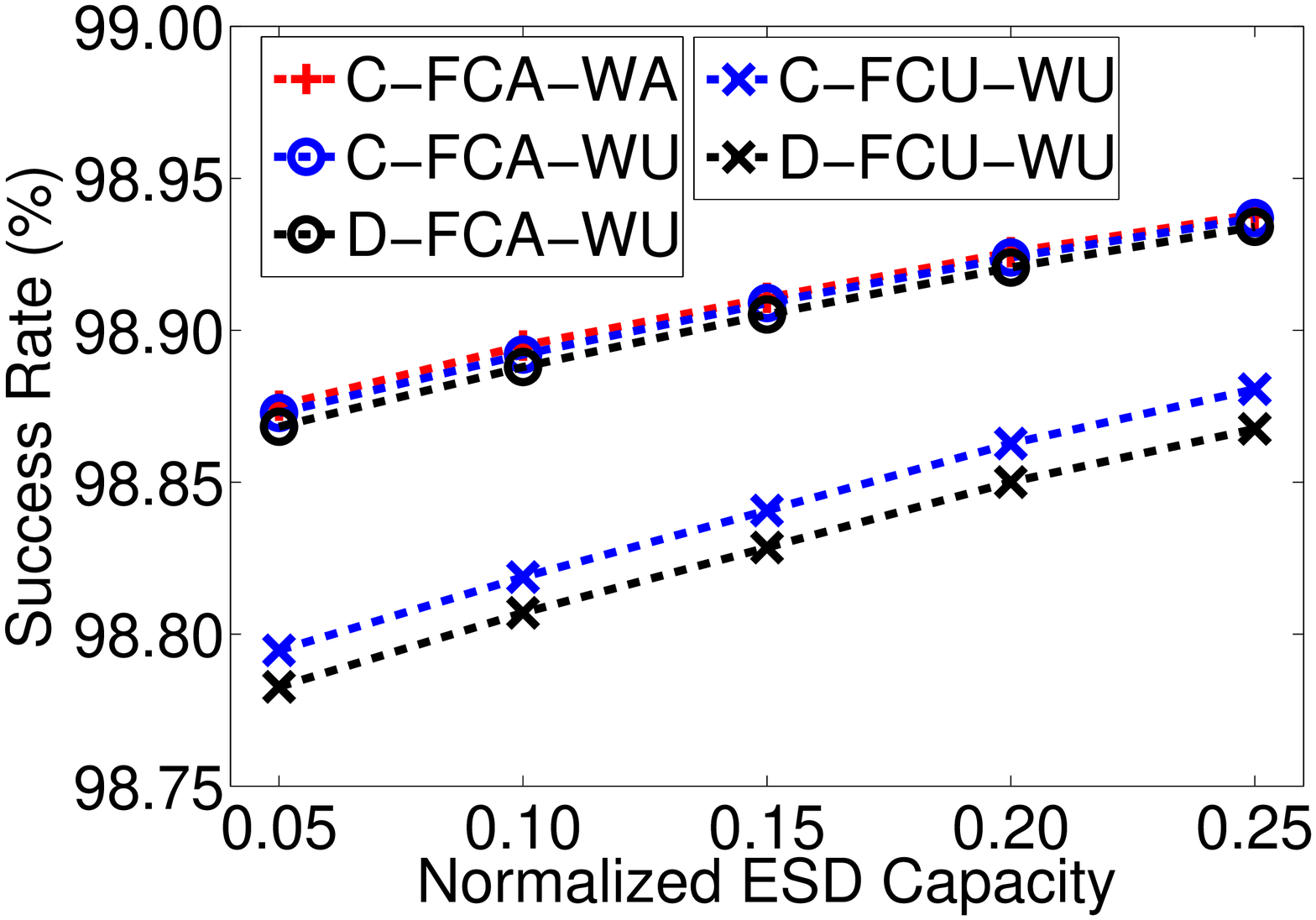}}\hfill{}\subfloat[Normalized average latency.]{\centering
\includegraphics[width=5.65cm]{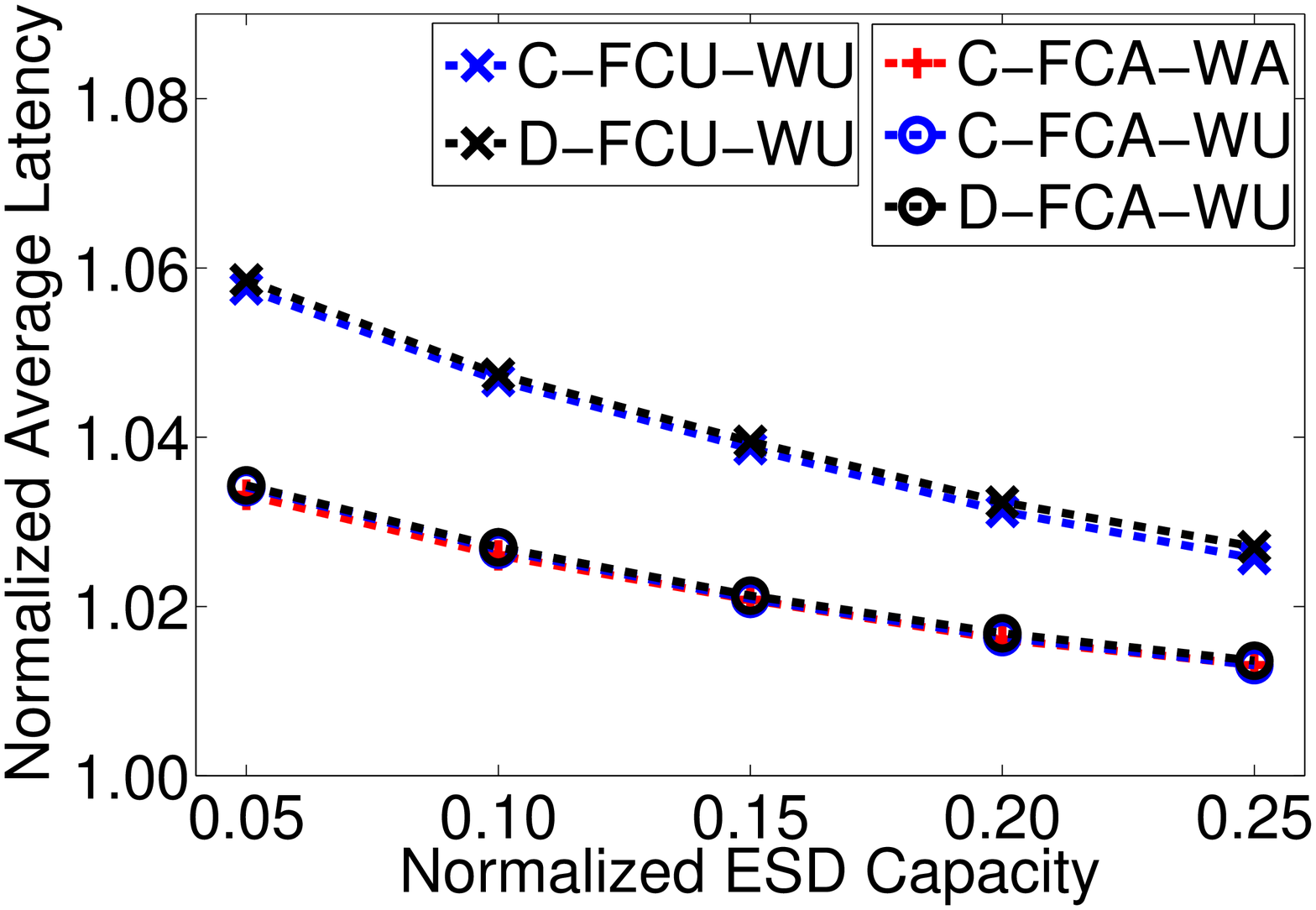}}\hfill{}\subfloat[Normalized
P95 latency.]{\centering
\includegraphics[width=5.65cm]{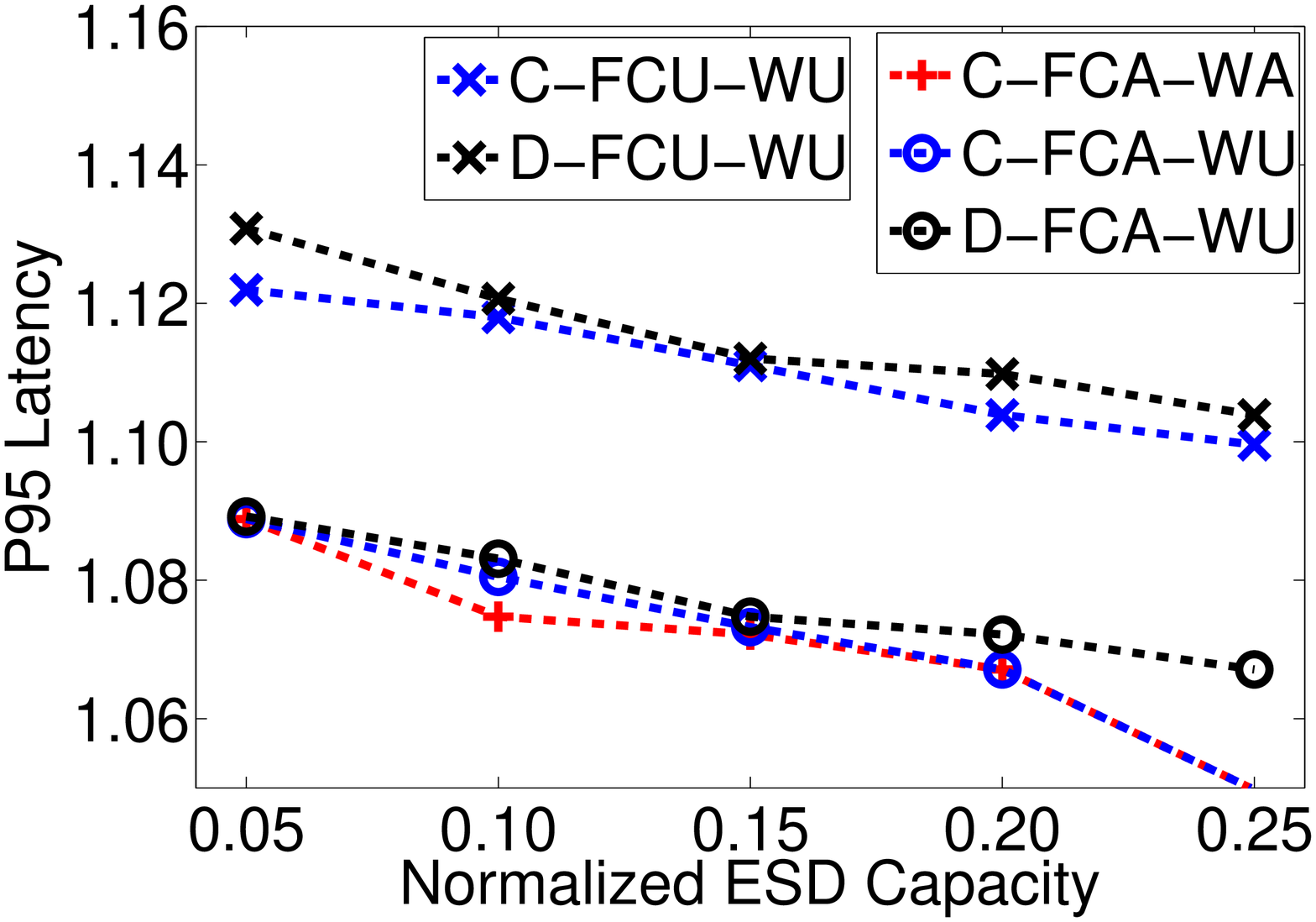}}
\vspace{-1pt}
\protect\caption{\label{fig:Trace1_performance}Workload performance of capping
policies under different ESD capacities for Trace~1.}
\end{figure*}

To understand the performance impact of our different power capping policies, we
study the success rate (Figure~\ref{fig:success-rate-capacity}) and
normalized average latency (Figure~\ref{fig:latency-capacity}) for the
synthesized workloads, sweeping over ESD capacity.  We observe that
as the ESD capacity decreases, the success rate decreases, while the average
latency increases.  
\changes{We also observe that more informed policies,
such as C-FCA-WA, can significantly reduce the performance degradation over more
oblivious policies such as D-FCU-WU.}
For example, at 50\% ESD capacity, D-FCU-WU has a
2.27\% drop in success rate and a 48\% increase in average latency, while
C-FCA-WA sees only a 0.64\% drop in success rate, and a 21\% average latency
increase.


We also explore how the average latency is impacted by changes in power surge
characteristics under our different power capping policies.  As before, we hold
all attributes constant except for the one being explored.  We examine latency
changes as a function of surge slope (Figure~\ref{fig:latency-slope}),
magnitude (Figure~\ref{fig:latency-magnitude}), and width 
(Figure~\ref{fig:latency-width}).  Again, we observe that smarter capping
policies that are centralized, or that are aware of fuel cell behavior or 
workload utility, are better at controlling the increase in latency for an ESD
underprovisioned at 50\%.  We also observe that for small slopes and small
magnitudes, the advantage of smarter policies becomes much smaller, as at
these smaller surge sizes, the underprovisioned ESD itself can handle the 
surge.  As a result, the capping policy is rarely invoked, and has little effect.
Interestingly, as Figure~\ref{fig:latency-width} shows, the latency initially
increases with the width, before dropping down and stabilizing.  This is
similar to the initial peak in latency that we observed in 
Figure~\ref{fig:Real-world-trace-analysis}c --- as the width grows, the
reliance on the power capping policy grows due to the underprovisioned ESD, but
over time, this reliance becomes steady, reducing the amount of capping required.

Finally, we ensure that our experiments are not sensitive to the period at which
\changes{we update the workload intensity heterogeneity between servers}.
As Figure~\ref{fig:latency-period} shows, the centralized policies maintain a constant
latency.  However, the decentralized policies drop slightly for longer update
periods, as the greater stability over the period allows the decentralized
heuristics, which use the starting power distribution between servers, to make
more accurate predictions.

\changes{We conclude that more intelligent power capping policies that take into account more information, such as centralized policies
or fuel cell aware and workload aware policies, are better able to utilize the
available ESD power when the ESD is underprovisioned, and thus deliver higher
performance.}

\begin{figure*}[t]
\vspace*{-15pt}
\subfloat[Success rate.]{\centering
\includegraphics[width=5.6cm]{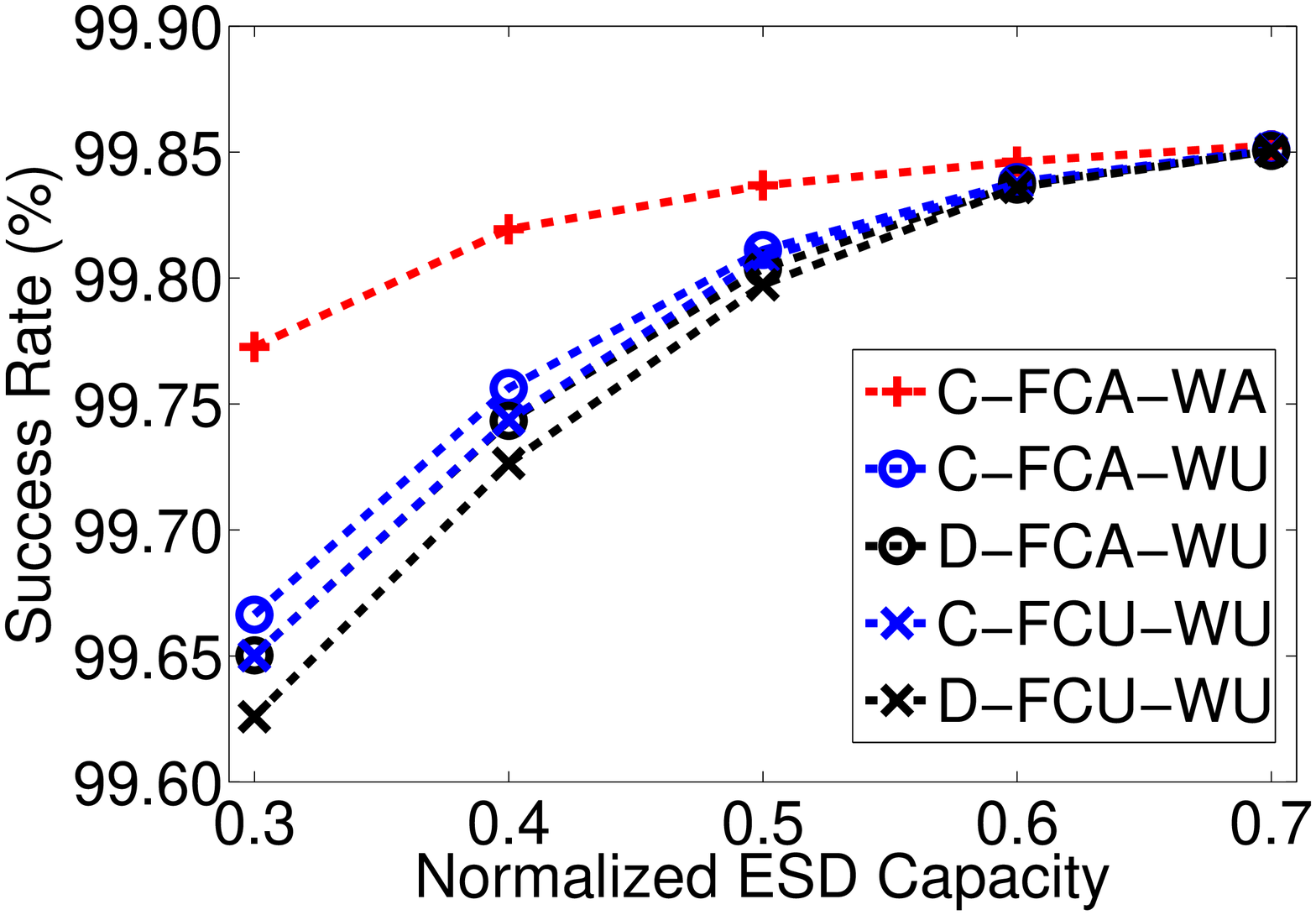}}\hfill{}\subfloat[Normalized average latency.]{\centering
\includegraphics[width=5.75cm]{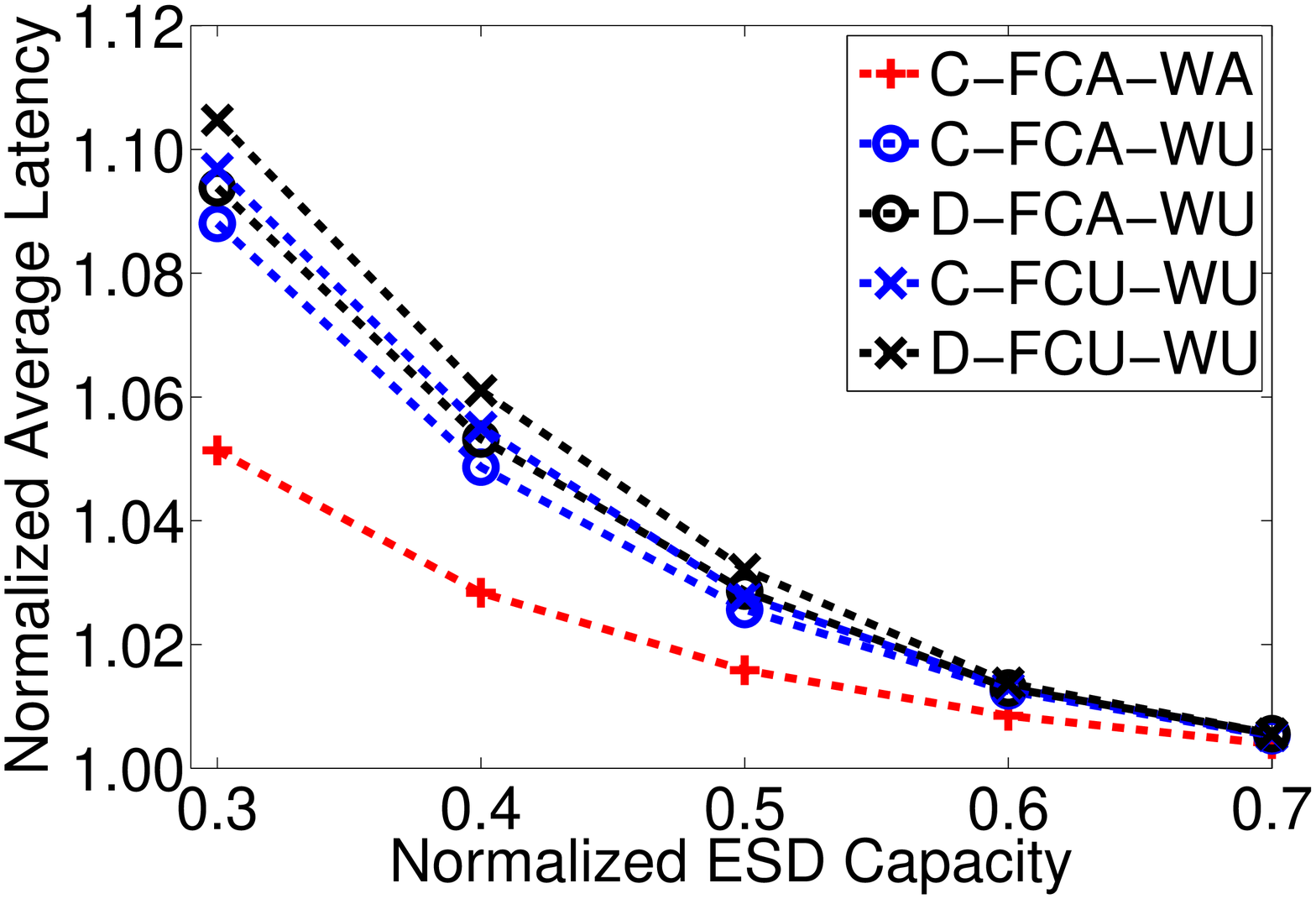}}\hfill{}\subfloat[Normalized
P95 latency.]{\centering
\includegraphics[width=5.6cm]{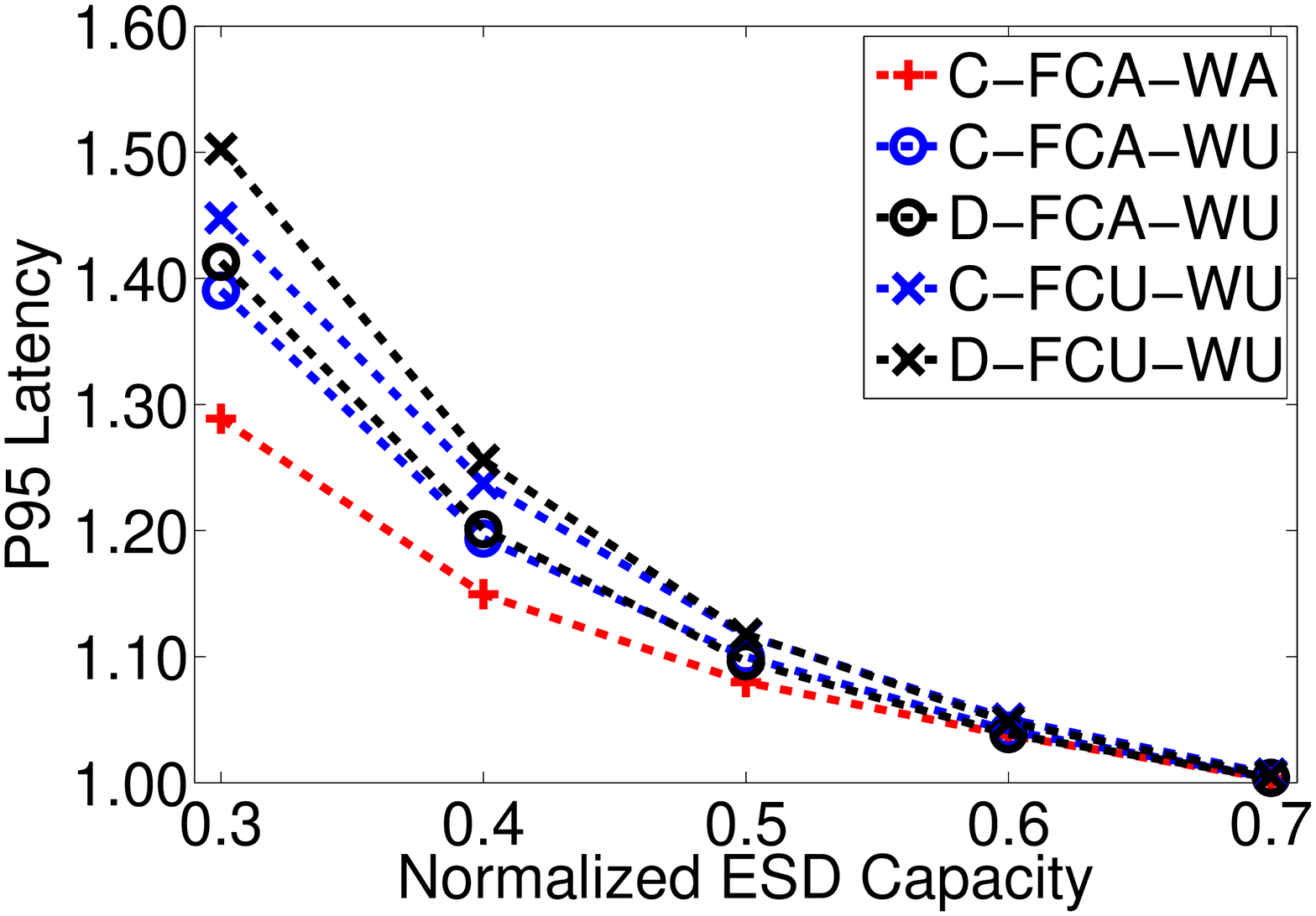}}
\vspace{-1pt}
\protect\caption{\label{fig:Trace2_performance}Workload performance of capping
 policies under different ESD capacities for Trace~2.}
\end{figure*}

\subsection{Production Data Center Traces}

We now evaluate the effectiveness of SizeCap using the two production data
center traces presented in Section~\ref{sec:analysis:trace}.  
Here, we employ the same method we used for our synthesized trace to generate the heterogeneity of workload intensity between servers.
Trace~1, shown in
Figure~\ref{fig:Real-world-trace-analysis}a, generally demands constant power,
but contains infrequent, large power surges.  Trace~2, shown in
Figure~\ref{fig:Real-world-trace-analysis}b, contains frequent power surges of
various sizes.  SizeCap must know the SLA of the
workload, as well as
what \changes{\emph{success rate margin} and \emph{average latency margin}}
the workload currently has to determine the
underprovisioned ESD \changes{capacity}. (For example, if a workload has an SLA of a 99.5\%
success rate, and its current success rate is 99.6\%, it has a success
rate margin of 0.1\%.)  For both traces, we assume the following margins
when they operate with a fully-provisioned ESD (i.e., an ESD that can tolerate
worst-case power surges without capping): 
0.1\% for success rate, 3\% for average latency, and 10\% for P95 latency.

Figure~\ref{fig:Trace1_performance} shows the
change in success rate, average latency, and P95 latency at various smaller
ESD sizes for Trace 1.  For both traces, with power
capping, as the ESD capacity decreases, workload performance reduces.  For 
Trace~1, which has little load power variation, we find that all of the fuel cell
aware policies perform similarly. If we look at the \mbox{D-FCA-WU} policy performance,
which is feasible to implement at scale in a contemporary data center, we find
that reducing the ESD capacity to \emph{as low as 15\%} of the fully-provisioned size
still meets our SLA requirements, with success rate reduction of only 0.1\%,\footnote{\changes{The success rates with a fully-provisioned ESD for Traces~1 and~2 are 99.00\% and 99.86\%, respectively.}}
an average latency increase of only 2.1\%, and a P95 latency increase of
7.5\%.

For Trace~2 (Figure~\ref{fig:Trace2_performance}), when the ESD capacity is
reduced, the workload performance degrades much faster than it does for Trace~1.
Unlike Trace~1, Trace~2 contains a large number of power surges, and therefore
invokes power capping much more frequently as the ESD capacity is reduced.
Looking at the D-FCA-WU policy, we find that we can still meet the SLA
requirements if we reduce the ESD capacity to as low as 50\% of the fully 
provisioned size.  At 50\% ESD capacity, the success rate reduces by only 
0.05\%,\changes{\footnotemark[\value{footnote}]}
the average latency increases by only 2.9\%, and the P95 latency increases 9.6\%.

For both traces, we leverage the life cycle model of a supercapacitor to 
evaluate the supercapacitor's lifetime. The model assumes that the supercapacitor cannot be used
beyond a certain number of charge/discharge cycles, and is based on recent
cycle testing of a commercial supercapacitor~\cite{murray2015cycle}. We find that under all of the ESD capacities
that we evaluated, for both traces, the ESD would have a lifetime of over
20~years.  As a result, we do not expect ESD lifetime to be a dominating
factor in data center TCO.

In summary, we find that SizeCap can significantly decrease the size of
the ESD for production data center workloads, without violating any SLA 
requirements.

\section{Related Work}
\label{sec:related}

\changes{
To our knowledge, this is the first work to (1)~holistically explore the
ESD sizing problem for fuel cell powered data centers (we use ESDs to
complement the power supply when the fuel cell is limited by its load
following behavior, which differs greatly from prior uses of ESD); and
(2)~develop power capping policies that are aware of load following,
and that can allow load following power sources such as fuel cells to continue 
ramping up power delivery during capping.
}




\vspace{3pt}
{\bf\noindent Energy Storage for Data Centers.}
ESDs in data centers have been mainly used for handling utility failures and/or intermittent power supplies such as 
renewable energy sources~\cite{wang.asplos14,Li2015ISCA,Sharma2011ASPLOS,Stewart2009HotPower, ricardo2010,goiri.asplos13,Li2012ISCA,Harsha_TPS}.
Recent studies~\cite{govindan.isca11,govindan.asplos12,DiWang_Sigmetrics2012,kontorinis.isca12,Li2015ISCA,urgaonkar.sigmetrics11} have proposed to leverage ESD-based peak shaving approaches 
to enhance data center demand response for cap-ex and op-ex savings. We leverage ESDs to complement 
fuel cell power supply load following limitations, which is very different from prior \changes{applications} of ESDs.


\vspace{3pt}
{\bf\noindent Power Capping for Data Centers.}
Many control algorithms for power capping have been proposed~\cite{femal.icac05,gandhi.sigmetrics09,wang.hpca08,wang.pact09,wang.tpds11,lim.atc11}.
More recent works have proposed power capping mechanisms
for data centers with renewable energy sources~\cite{li.hpca11,goiri.asplos13,ChaoLi_HPCA2013,ChaoLi_ICAC2014}, which leverage ESDs
to handle intermittent power losses.
All of these works assume that
ESDs have sufficient capacity, whereas our work aims to reduce the capacity of ESDs.

\vspace{3pt}
{\bf\noindent Fuel Cell Powered Data Centers.}
Riekstin et al.\ evaluate the TCO of using fuel
cells at the rack level~\cite{NoMoreElectrical}. Zhao et al.\
evaluate the energy efficiency benefits and load following capability
of fuel cells~\cite{LiZhao_TechnicalReport,Li_ESFuelCell}. While 
they consider ESDs, none of
these works explore the ESD sizing problem.


\section{Conclusion}
Fuel cells are a promising power source for data centers, but mechanical limitations
in fuel delivery make them slow to react to load surges, resulting in
\emph{power shortfalls}.  Prior work has used large ESDs to make up for such
shortfalls, but these large ESDs greatly increase the data center TCO.
 We analyzed the impact of various
power surge characteristics (slope, magnitude, and width) on ESD size,
and then demonstrated that power capping can effectively help to reduce the
size of the ESD required to cover shortfalls.  

Based on these observations, we \changes{propose} SizeCap, the first ESD sizing
framework for fuel cell powered data centers. Instead of using an ESD large
enough to cover the \emph{worst-case} load surges, SizeCap reduces the ESD size
such that it can handle most of the typical load surges, while still satisfying
SLA requirements. In the rare case when a worst-case surge occurs,
SizeCap employs power capping to ensure that the servers do not crash. As part
of our flexible framework, we \changes{propose} multiple power capping policies, each
with different degrees of awareness of fuel cell and workload behavior. Using
traces from \changes{Microsoft's} production data center systems, we \changes{show} that SizeCap can
successfully provide very large reductions in ESD size without violating any
SLAs. We conclude that SizeCap is a promising and low-cost framework for
managing power and TCO in fuel cell powered data centers.

\begin{figure*}[t]
\centering

\includegraphics[width=18cm]{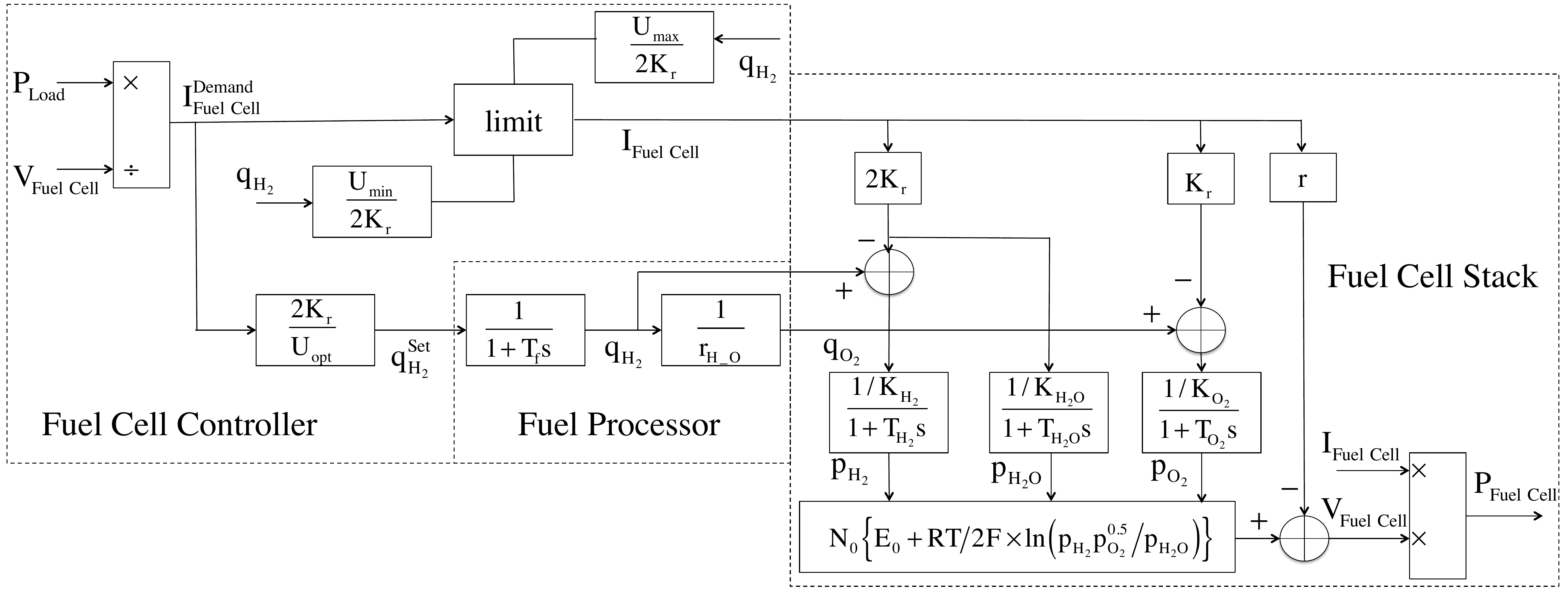}

\protect\caption{\label{fig:FuelCellModel}A typical model for fuel cell system.}

\end{figure*}

\begin{spacing}{0.9}
\vspace{-1pt}
\section*{Acknowledgments}
\vspace{-3pt}
We thank the anonymous reviewers from HPCA 2016 \changes{and members of the SAFARI research group}
for their constructive feedback. 
\changes{We thank Dr.\ Sanggyu Kang for helpful discussions on power system modeling.}
This work is supported by the NSF (award 1320531) and Microsoft Corporation. 
\vspace{-10pt}
\end{spacing}



\begin{spacing}{0.7}
\bstctlcite{bstctl:etal, bstctl:nodash, bstctl:simpurl}
\bibliographystyle{IEEEtranS}
\bibliography{references}
\end{spacing}
\section*{Appendix – Fuel Cell System Model}

\label{sec:Appendix}




We use a fuel cell system model based on \cite{padulles2000integrated,Fuel_Cell_Controller_Model,mufford1999power}, as Figure~\ref{fig:FuelCellModel} shows.
This model consists of three components – fuel cell controller model,
fuel processor model, and fuel cell stack model. 

\subsection*{Fuel Cell Controller Model}
The fuel cell controller model determines the fuel cell current $I_{Fuel\: Cell}$ and the set value for fuel flow rate $q_{H_{2}}^{Set}$ (here set value refers to the target to which the fuel flow rate is controlled).

First, the fuel cell controller model determines the demanded fuel cell current $I_{Fuel\, Cell}^{Demand}$
based on the load power $P_{Load}$
and fuel cell voltage $V_{Fuel\, Cell}$:

\begin{equation}
I_{Fuel\, Cell}^{Demand} = P_{Load} / V_{Fuel\, Cell}
\end{equation}

In the meanwhile, the fuel cell controller model calculates the range of fuel cell current that the fuel cell stack can safely provide:

\begin{equation}
\frac{U_{min}}{2K_{r}}q_{H_{2}}<I_{Fuel\, Cell}<\frac{U_{max}}{2K_{r}}q_{H_{2}}
\end{equation}

\noindent Where $q_{H_{2}}$ is the fuel flow rate; $2K_{r}$ is the conversion coefficient between fuel cell current and fuel flow rate. Dividing the fuel flow rate by the conversion coefficient, we can get the fuel cell current corresponding to the fuel flow rate assuming 100\% fuel utilization. However, in reality, a fuel cell stack can safely operate only within a certain range of fuel utilization rate. Hence, by further multiplying with the maximum/minimum safe fuel utilization rate ($U_{max}$ / $U_{min}$), we can get the range of fuel cell current that the fuel cell stack can safely provide. 

Then, the fuel cell controller model compares the demanded current $I_{Fuel\, Cell}^{Demand}$ with the range of current that the fuel cell stack can safely provide, and determines the fuel cell current $I_{Fuel\: Cell}$.

Besides determining the fuel cell current, the fuel cell controller also performs feedforward control on
the fuel flow rate, by estimating the fuel flow rate that can satisfy
the load power as the set value for fuel flow rate $q_{H_{2}}^{Set}$:

\begin{equation}
q_{H_{2}}^{Set}=\frac{2K_{r}}{U_{opt}}I_{Fuel\, Cell}^{Demand}
\end{equation}

\noindent Where $U_{opt}$ is the optimal fuel utilization rate that is best for fuel cell stack health and operation.

\subsection*{Fuel Processor Model}
The fuel processor model computes the fuel flow rate $q_{H_{2}}$ and the corresponding oxygen flow rate $q_{O_{2}}$.  

Based on the set value for fuel flow rate, the fuel processor gradually
adjusts its fuel flow rate $q_{H_{2}}$
and oxygen flow rate $q_{O_{2}}$.
As prior work points out, due to the mechanical nature of fuel cells, fuel flow
rate cannot be immediately adjusted to its set value, and this delay
is the root cause of the fuel cell load following limitations \cite{mueller2009intrinsic,Li_ESFuelCell}.
Zhu and Tomsovic~\cite{Fuel_Cell_Controller_Model} modeled this delay using a first-order dynamic system: 

\begin{equation}
q_{H_{2}}\left(s\right)=q_{H_{2}}^{Set}\left(s\right)\times\frac{1}{1+T_{f}s}
\end{equation}
\begin{equation}
q_{O_{2}}=q_{H_{2}}\frac{1}{r_{H\_ O}}
\end{equation}

\noindent Where $q_{H_{2}}\left(s\right)$ and $q_{H_{2}}^{Set}\left(s\right)$ are the Laplace transforms of fuel flow rate and its set value; $T_{f}$ is the time constant for this dynamic system; $r_{H\_ O}$ is the ratio between fuel flow rate and oxygen flow rate.

\subsection*{Fuel Cell Stack Model}

Fuel cell stack model computes the fuel cell output power $P_{Fuel\, Cell}$ based on the fuel cell current $I_{Fuel\, Cell}$, fuel flow rate $q_{H_{2}}$, and oxygen flow rate $q_{O_{2}}$.

To do this, first, the fuel cell stack model needs to determine the partial pressure of fuel, oxygen, and water ($p_{H_{2}}$, $p_{O_{2}}$, $p_{H_{2}O}$) in the fuel cell stack. The partial pressure of each component (fuel, oxygen, and water) is dependent on the amount of that component in the fuel cell stack, which is further impacted by the input flow rate, depletion rate (the rate that the component is consumed by the fuel cell stack, which is proportional to the fuel cell current $I_{Fuel\, Cell}$), and output flow rate (the rate that the compoent flows out of the fuel cell stack; Padulles et al.~\cite{padulles2000integrated} found it can be modeled as being proportional to the partial pressure of the component). Through sophisticated modeling and derivation, Padulles et al.~\cite{padulles2000integrated} found we can use three first-order dynamic systems to model how the partial pressure of fuel, oxygen, and water evolves across time:

\begin{equation}
p_{H_{2}}\left(s\right)=\frac{1/K_{H_{2}}}{1+T_{H_{2}}s}\left(q_{H_{2}}\left(s\right)-2K_{r}I_{Fuel\, Cell}\left(s\right)\right)
\end{equation}

\begin{equation}
p_{O_{2}}\left(s\right)=\frac{1/K_{O_{2}}}{1+T_{O_{2}}s}\left(q_{O_{2}}\left(s\right)-K_{r}I_{Fuel\, Cell}\left(s\right)\right)
\end{equation}

\begin{equation}
p_{H_{2}O}\left(s\right)=\frac{1/K_{H_{2}O}}{1+T_{H_{2}O}s}\left(-2K_{r}I_{Fuel\, Cell}\left(s\right)\right)
\end{equation}

\noindent Where $p_{H_{2}}\left(s\right)$, $p_{O_{2}}\left(s\right)$, and $p_{H_{2}O}\left(s\right)$ are the Laplace transforms of partial pressure of fuel, oxygen, and water; $q_{H_{2}}\left(s\right)$ and $q_{O_{2}}\left(s\right)$ are the Laplace transforms of the fuel flow rate and oxygen flow rate; $I_{Fuel\, Cell}\left(s\right)$ is the Laplace transform of the fuel cell current; $K_{H_{2}}$, $K_{O_{2}}$, and $K_{H_{2}O}$ are the valve molar constants for fuel, oxygen, and water; $T_{H_{2}}$, $T_{O_{2}}$, and $T_{H_{2}O}$ are the time constants of the dynamic systems corresponding to fuel, oxygen, and water.

After getting the partial pressure of each component, we can get the fuel cell voltage $V_{Fuel\, Cell}$ by applying Nernst's equation and Ohm's law (to consider ohmic losses)~\cite{padulles2000integrated}:

\begin{equation}
V_{Fuel\, Cell}=N_{0}\left\{ E_{0}+\frac{RT}{2F}\times\ln\left(\frac{p_{H_{2}}p_{O_{2}}^{0.5}}{p_{H_{2}O}}\right)\right\} -rI_{Fuel\, Cell}
\end{equation}

\noindent Where $N_{0}$ is the number of fuel cells connected in series in the fuel cell stack; $E_{0}$ is the voltage associated with the reaction free energy for fuel cells; $R$ and $F$ are the universal gas constant and Faraday's constant respectively; $T$ is the fuel cell stack temperature; $r$ describes the ohmic losses of the fuel cell stack.

Finally, by multiplying the fuel cell voltage with the fuel cell current, we can obtain the fuel cell output power as:
\begin{equation}
P_{Fuel\, Cell}=V_{Fuel\, Cell}\times I_{Fuel\, Cell}
\end{equation}

In this way, we build the fuel cell system model by connecting the fuel cell controller model, the fuel processor model, and the fuel cell stack model. This fuel cell system model can compute the fuel cell (output) power trace $\left\{ P_{Fuel\, Cell}\left(t\right)\right\} $
under an arbitrary load power trace $\left\{ P_{Load}\left(t\right)\right\} $, which is used in this paper.

\end{document}